\documentclass[iop, apj, numberedappendix, twocolappendix]{emulateapj}

\usepackage{hyperref}
\usepackage{xcolor}
\usepackage{amsmath}
\definecolor{dark-red}{rgb}{0.4,0.15,0.15}
\definecolor{dark-blue}{rgb}{0.15,0.15,0.4}
\definecolor{medium-blue}{rgb}{0,0,0.5}
\hypersetup{
    colorlinks,
    linkcolor={black},
    citecolor={black},
    urlcolor={black}
}
\usepackage[varg]{txfonts}

\newcommand{\para}{\parallel}

\newcommand{\paren}[1]{ \left( #1 \right) }
\newcommand{\kb}{k_{\rm B}}

\newcommand{\kz}{k_z (\lambda_{\rm mfp} H)^{1/2}}

\newcommand{\kx}{k_x (\lambda_{\rm mfp} H)^{1/2}}
\newcommand{\dmu}{d\ln \mu/d\ln P}
\newcommand{\dT}{d\ln T/d\ln P}
\renewcommand{\b}{\hat{\bb{b}}}

\newcommand{\be}{\begin{eqnarray}}
\newcommand{\en}{\end{eqnarray}}
\newcommand{\pa}{\partial}
\newcommand{\f}{\frac}
\newcommand{\omc}{\omega_{\rm c}}
\newcommand{\omd}{\omega_{\rm D}}
\newcommand{\omv}{\omega_{\rm v}}
\newcommand{\oma}{\omega_{\rm A}}

\newcommand{\omdy}{\omega_{\rm dyn}}

\newcommand{\D}[2]{\frac{\partial #1}{\partial #2}}
\newcommand\bb[1]{\mbox{\boldmath{$#1$}}}
\newcommand\del{\bb{\nabla}}
\newcommand\bcdot{\bb{\cdot}}
\newcommand\btimes{\bb{\times}}
\newcommand{\unitspace}{\ensuremath{\,}}
\newcommand{\usp}{\unitspace}

\usepackage{color}
\definecolor{brown}{rgb}{0.42,0.24,0.07}
\definecolor{darkgreen}{rgb}{0.0,0.6,0.00}
\definecolor{purple}{rgb}{0.7,0.0,0.7}

\begin{document}

\title{Plasma Instabilities in the Context of Current Helium Sedimentation Models: \\
Dynamical Implications for the ICM in Galaxy Clusters}

\author{Thomas Berlok and Martin E. Pessah}
\affil{Niels Bohr International Academy, Niels Bohr Institute, Blegdamsvej 17, 2100 Copenhagen, Denmark}
\email{berlok@nbi.dk}

\shorttitle{Instabilities in Weakly-Collisional Plasmas}
\shortauthors{Berlok \& Pessah}

\begin{abstract}
Understanding whether Helium can sediment to the core of galaxy clusters is important for a number of problems in cosmology and astrophysics. All current models addressing this question are one-dimensional and do not account for the fact that magnetic fields can effectively channel ions and electrons, leading to anisotropic transport of momentum, heat, and particle diffusion in the weakly collisional intracluster medium (ICM). This anisotropy can lead to a wide variety of instabilities, which could be relevant for understanding the dynamics of heterogeneous media. In this paper, we consider the radial temperature and composition profiles as obtained from a state-of-the-art Helium sedimentation model and analyze its stability properties. We find that the associated radial profiles are unstable, to different kinds of instabilities depending on the magnetic field orientation, at all radii. The fastest growing modes are usually related to generalizations of the Magnetothermal Instability (MTI) and the Heat-flux-driven Buoyancy Instability (HBI) which operate in heterogeneous media. We find that the effect of sedimentation is to increase (decrease) the predicted growth rates in the inner (outer) cluster region. The unstable modes grow fast compared to the sedimentation timescale. This suggests that the composition gradients as inferred from sedimentation models, which do not fully account for the anisotropic character of the weakly collisional environment, might not be very robust. Our results emphasize the subtleties involved in understanding the gas dynamics of the ICM and argue for the need of a comprehensive approach to address the issue of Helium sedimentation beyond current models.
\end{abstract}


\section{Introduction}
Galaxy clusters are important astrophysical probes since their masses can be used to constrain
cosmological parameters (\citealt{2014MNRAS.440.2077M} and references therein).
The distribution of mass as a function of radius
can be inferred by modeling the observed X-ray emission produced by Bremsstrahlung in the
hot intracluster medium (ICM).
The intensity of the emission depends on the radial distribution of temperature and density, as well as the
composition of the gas. While the temperature of the ICM is reasonably well determined \citep{Vik06}, the
composition of the plasma  is not. The reason is that many of the elements are completely
ionized at the characteristic temperatures of the ICM and thus their abundances cannot be
directly inferred.  Therefore, the interpretation of the X-ray data normally relies on assuming
a model for the  composition of the gas. A widely adopted approximation consists on assuming
the composition of the plasma to be uniform (see \citealt{Bul11} for an analysis where this assumption
is relaxed).
Elements heavier than hydrogen are expected to sediment over cosmological timescales \citep{Fab77},
therefore the assumption of a homogeneous ICM relies on this
process being inefficient. Turbulence and tangled magnetic fields, or a combination of both,
have been invoked as potential agents \citep{Mar07}.

Even though the mass ratio between Helium (He) and Hydrogen (H) is small, because He is the most abundant of
the heavy elements, it has the potential to induce significant variations in the mean molecular
weight. If He sedimentation does take place and this is not accounted for when modeling galaxy clusters,
this could induce biases in the cosmological parameters derived \citep{Bo00,Mar07,Pen09}.
This could prove to be a problem for precision cosmology and highlights the importance of
understanding  the distribution of heavy elements in the ICM \citep{Fab77,Gil84,Chu03,Chu04,Pen09,Sht10}.
Most of the previous work on this subject is  based on solving Burgers' equations for a multicomponent plasma
\citep{1969fecg.book.....B,Tho94} and all of these assume spherical symmetry in order to predict  the composition of the ICM as a function of radius. Studies addressing the long term evolution of the composition of the ICM have considered the dynamical effects of magnetic fields in a rather crude way, usually encapsulating their effects in a parameter that regulates the slow down
of the sedimentation process \citep{Pen09}.

A more recent,  and somewhat parallel, line of developments has helped us realize that the dynamical properties of magnetized, weakly collisional, stratified plasmas can be rather subtle. \citet{Bal00,Bal01} and \citet{Qua08} showed that stratified plasmas that are stable according to the Schwarzschild criterion could turn unstable due to the presence of a magnetic field, even if its strength is too weak to be
mechanically important. The plasma can become unstable because even a  very weak magnetic field can effectively alter transport processes by channeling electrons and ions, leading to anisotropic heat conduction and Braginskii viscosity \citep{Bra,Kun11}.

Previous studies have considered plane-parallel, fully ionized \emph{homogeneous} atmospheres with a temperature gradient in the direction of gravity. In this setting there are two instabilities that feed on the gradient in temperature.
The Magnetothermal Instability (MTI) has the fastest growth rate when the magnetic field is perpendicular to gravity and the temperature decreases with height \citep{Bal00,Bal01}.
The Heat-flux-driven Buoyancy Instability (HBI) has the fastest growth rate when the magnetic field is parallel to gravity and the temperature increases with height \citep{Qua08}.
Because of the temperature profiles observed in typical cool-core galaxy clusters \citep{Vik06}, the MTI is believed to be active in the outer parts of the ICM  while the HBI is believed to be relevant in the inner parts of the ICM. These instabilities have been studied extensively in the literature both analytically \citep{Bal00,Bal01,Qua08,Kun11, Lat12} and numerically with initially local simulations with anisotropic heat conduction \citep{Par05,Par07,Par08} and since then with elaborate physical models \citep{Par08_MTI,Par09,Bog09,2010ApJ...712L.194P,2010ApJ...713.1332R,2011MNRAS.413.1295M,2012MNRAS.419.3319M,Kun12,Par12,2012MNRAS.419L..29P}.

The aforementioned works that deal with the weakly collisional character of the magnetized plasma
have usually adopted a homogeneous atmosphere as a model for the ICM.
On the other hand, the sedimentation models are usually one-dimensional and do not fully account for
dynamical properties of the magnetic field.
In an effort to better understand the interplay between the Helium distribution in the ICM and its weakly collisional and weakly magnetized nature, \citet{Pes13} considered the presence of a gradient in the Helium composition and extended previous stability criteria. Their work shows that a gradient in composition can modify the stability properties of a stratified atmosphere. This could have consequences for the Helium sedimentation models which could be unstable to plasma instabilities.

The equations used to model the plasma in \cite{Pes13} describe the stability properties of a weakly collisional plasma subject to a background composition gradient, but they do not account for the process of Helium sedimentation which is estimated to occur on longer timescales.\footnote{We discuss the limitations of this work in this regard in Appendix \ref{sec:sedi_timescales}.} Because of this, the equations are thus unable to predict how a gradient in composition arises from an initial homogeneous plasma.
{A framework that simultaneously considers the physics responsible for Helium sedimentation together with
the anisotropic transport properties governing dilute, magnetized plasmas has yet to be developed.
A key goal for the future is therefore to develop such a model in order to determine from first principles the rate at which Helium can sediment in a weakly collisional, magnetized medium.
In lieu of such a fully consistent theory, this paper has a more modest goal. Our aim is to understand the kind of
instabilities, and their associated timescales and length scales, that can feed off the temperature and composition
profiles that emerge from state-of-the-art models for Helium sedimentation in the ICM \citep{Pen09,Sht10}.

The rest of the paper is organized as follows.
In Section 2, we introduce the equations that we use to model the weakly collisional three-component plasma.
In Section 3, we derive an extended version of the dispersion relation presented by \cite{Pes13} to account for the effects of magnetic tension, which can be important in cluster cores.
In Section 4, we discuss the stability criteria for atmospheres with temperature and composition gradients.
In Section 5, we solve the dispersion relation for isothermal atmospheres in order to gain insight into the
type of instabilities that can be excited solely by composition gradients.
In Section 6, we consider the temperature and composition gradients derived from
the Helium sedimentation model of \cite{Pen09}. By allowing the background magnetic field to have an arbitrary inclination with respect to gravity we identify the most relevant instabilities in different regions of the ICM.
Finally, we conclude by discussing  future prospects for addressing the problem of Helium sedimentation
in galaxy clusters on more fundamental grounds in Section 7.

\section{The equations of kinetic MHD for a binary mixture}

The kinetic MHD equations for a fully ionized binary mixture of Hydrogen and Helium
can be written as \citep{Pes13}
\be &&\f{\pa \rho}{\pa t}+\bb{\nabla}\bcdot(\rho \bb{v})=0 \,,
\label{eq:rho}\\
&&\D{\paren{\rho \bb{v}}}{t}
+\del \bcdot \paren{\rho \bb{vv} + P_\mathrm{T} \mathbf{I} - \frac{B^2}{4\pi} \b\b}
=
-\del \bcdot \Pi
+ \rho \bb{g}  , \, \\
&&\f{\pa \bb{B}}{\pa t}=\bb{\nabla}\btimes(\bb{v}\btimes\bb{B}) \,,
\label{eq:b}\\
&&
\f{\rho T}{\mu} \f{ds}{dt}
 = -\del \bcdot \bb{Q}_{\rm s} - \Pi \bb{:} \del \bb{v}
 \,, \,\,\,\,\,\,\,
\label{eq:S} \\
&&\f{dc}{dt} =-\bb{\nabla}\bcdot\bb{Q}_{\rm c} \,.
\label{eq:c}
\en
Here, the Lagrangian and Eulerian derivatives are related via
$d/dt = \pa/\pa t + \bb{v} \bcdot \nabla$, $\rho$ is the mass
density, $\bb{v}$ is the fluid velocity, $\bb{g} = (0,
0, -g)$ is the gravitational
acceleration and $\mathbf{I}$ stands
for the $3 \times 3$ identity matrix.  The symbols $\bot$ and
$\parallel$ refer respectively to the directions perpendicular and parallel to the
magnetic field $\bb{B}$ whose
direction is given by the unit vector $\hat{\bb{b}} = \bb{B}/B = (b_x,
0, b_z)$. The total pressure is $P_\mathrm{T} = P+ {B^2}/{8	\pi}$, where
$P$ is the thermal pressure, and
the entropy per unit mass is defined by
\be
s = \f{3\kb}{2 m_{\rm{H}}} \ln \paren{P\rho^{-5/3}} \ ,
\en
where $\kb$ is Boltzmann's constant and $m_{\rm H}$ is the proton mass.
The adiabatic index, $\gamma$, has been set to $5/3$ in the preceding equations and
throughout the remainder of the paper.

The composition of the plasma, $c$,
is defined to be the ratio of the Helium density
to the total gas density
\be
c = \f{\rho_{\rm He}}{\rho_{\rm H} + \rho_{\rm He}} \ .
\en
The associated mean molecular weight, $\mu$, influences
the dynamics of the plasma through the equation of state
\be
P= \frac{\rho k_{\rm B} T}{\mu m_{\rm H}} \, , \label{eq:eos}
\en
where $T$ is the temperature.
We assume a completely ionized plasma consisting of Helium and Hydrogen
and the mass concentration of Helium, $c$, is therefore related to the mean molecular
weight, $\mu$, by
\begin{equation}
\mu=\frac{4}{8-5c} \ .
\end{equation}
The evolution of the binary mixture is influenced by three different non-ideal effects, namely
Braginskii viscosity, which is described through the viscosity tensor \citep{Bra}
\be
\Pi = - 3\rho \nu_\para \paren{\b\b -\f{1}{3} \mathbf{I} } \paren{ \b\b -\frac{1}{3}\mathbf{I} }
 \bb{:} \del \bb{v}  \ ,
\en
anisotropic heat conduction described by the heat flux \citep{1962pfig.book.....S,Bra}
\begin{equation}
\bb{Q}_{s}=-\chi_\para\hat{\bb{b}}\hat{\bb{b}}\cdot\nabla T,
\end{equation}
and  anisotropic diffusion of Helium described by the composition flux
\begin{equation}
\bb{Q}_{c}=-D\hat{\bb{b}}\hat{\bb{b}}\cdot\nabla c \ .
\end{equation}
The transport coefficients ($\chi_\para$, $\nu_{\para}$, and $D$) all depend on the temperature, as well as the composition of the plasma. The dependences are given in Appendix \ref{sec:transport_properties} by Equations (\ref{eq:chi_para}), (\ref{eq:eta_para}) and
(\ref{eq:Bahcael_dif}), respectively. For more details on the kinetic MHD approximation and its limitations
see the relevant discussions in \citet{schekochihin_plasma_2005}, \citet{Kun12}, \citet{Pes13} and references therein.

\section{The dispersion relation}
We consider an initially motionless, plane-parallel atmosphere with gradients in both temperature and the mean molecular weight.
A local linear mode analysis of this atmosphere, using Equations (\ref{eq:rho})--(\ref{eq:c}) and following the procedure in \cite{Pes13},
leads to the dispersion relation
\be
\sum_{i=0}^{4}A_{i}+\omega_{\rm v}\sum_{i=1}^{5}B_{i}=0 \label{eq:disp_relation} \ ,
\en
where the coefficients are given by
\be
A_{0} & = & \sigma^{2}\tilde{\sigma}^{4}k^{2} \ , \\
A_{1} & = & \sigma\tilde{\sigma}^{4}\left(\omega_{D}+\omega_{c}\right)k^{2}\ ,\\
A_{2} & = & \sigma^{2}\tilde{\sigma}^{2}N^{2}\left(k_{x}^{2}+k_{y}^{2}\right)+\tilde{\sigma}^{4}\omega_{c}\omega_{D}k^{2}\ ,\\
A_{3} & = & \sigma\tilde{\sigma}^{2}g\omega_{c}\left\{ \frac{d\ln T}{dz}\mathcal{K}-\frac{d\ln\mu}{dz}\left(k_{x}^{2}+k_{y}^{2}\right)\right\} \nonumber \\
&& +\sigma\tilde{\sigma}^{2}\omega_{D}N^{2}\left(k_{x}^{2}+k_{y}^{2}\right)\ ,\\
A_{4} & = & \tilde{\sigma}^{2}\omega_{c}\omega_{D}N_{T/\mu}^{2}\mathcal{K}\ , \\
B_{1} & = & \sigma^{3}\tilde{\sigma}^{2}k_{\perp}^{2}\ , \\
B_{2} & = & \sigma^{2}\tilde{\sigma}^{2}\left(\omega_{D}+\omega_{c}\right)k_{\perp}^{2}\ , \\
B_{3} & = & \sigma^{3}b_{x}^{2}k_{y}^{2}N^{2}+\sigma\tilde{\sigma}^{2}\omega_{c}\omega_{D}k_{\perp}^{2}\ , \\
B_{4} & = & \sigma^{2}b_{x}^{2}k_{y}^{2}\left(N_{T/\mu}^{2}\omega_{c}+N^{2}\omega_{D}\right)\ , \\
B_{5} & = & \sigma b_{x}^{2}k_{y}^{2}N_{T/\mu}^{2}\omega_{c}\omega_{D} \ .
\en
Here, $k_\parallel = \b \bcdot \bb{k}$ and $k_\perp^2 = k^2 - k_\parallel^2$ and we have defined
\be
\omega_{\rm c} = \f{2}{5} \,\f{\chi_\para T}{P} k_{\parallel}^{2} \ , \label{eq:omega_c} \quad
\omega_{\rm v} = 3\nu_{\parallel} k_{\parallel}^{2}\ , \quad
\omega_{\rm D} = D k_{\parallel}^{2}\ , \quad \; \;
\en
which are the inverse timescales associated with anisotropic heat conduction, viscosity and
particle diffusion.
Furthermore, we have introduced the quantity
\be
\mathcal{K}=\left(1-2b_{z}^{2}\right)\left(k_{x}^{2}+k_{y}^{2}\right)+2b_{x}b_{z}k_{x}k_{z} \,,
\label{eq:geometric_curly_K}
\en
as well as the Brunt$-$V{\"a}is{\"a}l{\"a}  frequency, $N$, such that
\be
N^2 = \frac{2}{5}\f{m_{\rm{H}}}{\kb}g\frac{ds}{dz} \ ,
\en
and the quantity
\be
N_{T/\mu}^2 = g\frac{d}{dz}\ln\left(\frac{T}{\mu}\right) \,.
\en
The effects of magnetic tension, which are neglected in \cite{Pes13} and
could be important in the inner parts of the ICM \citep{Car02}, are contained in
\be
\tilde{\sigma}^2 = \sigma^2+\oma^{2} \ ,
\en
where
\be
\omega_{\rm A} = k_{\parallel}v_{\rm A} \ ,\label{eq:omega_A}
\en
is the Alfv{\'e}n frequency and $v_{\rm A}= {B}/\sqrt{4 \pi \rho}$ is the Alfv{\'e}n velocity.

\subsection{Characteristic Scales and Dimensionless Variables}

There are a number of characteristic scales that are useful to introduce.
The dynamical frequency, $\omega_{\rm dyn}$, is given by
\be
\omega_{\rm dyn} &=& \sqrt{\f{g}{H}} \label{eq:omega_dyn} \ ,
\en
where $H$ is the thermal pressure scale height and $g$ is the gravitational acceleration.
We will use that hydrostatic equilibrium requires
\be
g \f{d\ln P}{dz} = -\omdy^2 \ .
\en
The plasma-$\beta$, given by the ratio of the thermal velocity and
the Alfv{\'e}n speed squared
$
\beta  =  {v_{\rm th}^{2}}/{v_{\rm A}^{2}},
$
where
$
v_{\rm th}^2 = P/\rho
$,
provides a measure of the strength of the magnetic field.

We also define the Knudsen number
\be
\textrm{Kn} = \f{\lambda_{\rm mfp}}{H}\ ,
\en
which is a measure of the collisionality of the plasma.
Here, $\lambda_{\rm mfp}$ is the mean-free-path of ion collisions. Intuitively,  $\textrm{Kn}^{-1} = H/\lambda_{\rm mfp}$ is the average number of collisions
an ion experiences as it traverses a distance of one scale height. So $\textrm{Kn}^{-1} \gg 1$
($\textrm{Kn}^{-1} \ll 1$) corresponds to high (low) collisionality.
As in \cite{Pes13}, we define an effective ion-ion collision frequency
\be
\nu^{\rm eff}_{ii} = \f{v_{\rm th}^2}{2 \nu_\para} \ ,
\en
which can be used to express the inverse Knudsen number as
\be
\textrm{Kn}^{-1} = \f{\nu^{\rm eff}_{ii}}{\omega_{\rm dyn}} \ ,
\en
by using that $\lambda_{\rm mfp} = v_{\rm th}/\nu^{\rm eff}_{ii}$.

In Section \ref{sec:isothermal}, it will prove useful to use a dimensionless form
of the theory  and present the results in terms of variables that have been scaled
using the characteristic time provided by $\omdy^{-1}$ and the characteristic
length given by $(\lambda_{\rm mfp} H)^{1/2}$. In order to accomplish this, we
assume that the inverse timescales for heat conduction and Braginskii
viscosity are related to the dynamical frequency via \citep{Kun11}
\be
\omc & \simeq &  10 k_\parallel^2 \lambda_{\rm mfp} H  \omdy \,, \label{eq:conduction_time} \\
\omv & \simeq & \f{3}{2} k_\parallel^2 \lambda_{\rm mfp} H \, \omdy \, . \label{eq:viscosity_time}
\en
When diffusion of Helium is included in the analysis we furthermore assume that
\be
\omd & \simeq &  \f{1}{4} k_\parallel^2 \lambda_{\rm mfp} H \, \omdy \,. \label{eq:diffusion_time}
\en
The approximations given by Equations (\ref{eq:conduction_time})-(\ref{eq:diffusion_time}) are justified in Appendix \ref{sec:time_scale_for_diffusion}.

Note that the local linear analysis leading to the dispersion relation in
\cite{Pes13} is only valid when the wavenumbers involved
satisfy the inequalities
\be
\sqrt{\textrm{Kn}} &\ll& k_\para (\lambda_{\rm mfp} H)^{1/2} \ll \sqrt{\textrm{Kn}^{-1}} \label{eq:fluid_limit}\ , \\
\f{1}{10 \sqrt{\beta\textrm{Kn}}} &\ll& k_\para (\lambda_{\rm mfp} H)^{1/2} \ll \sqrt{\beta\textrm{Kn}} \label{eq:magnetic_tension}\ .
\en
The dispersion relation in Equation (\ref{eq:disp_relation}) is also valid even when the inequality given by Equation (\ref{eq:magnetic_tension}) is not fulfilled because the effects of magnetic tension,
which are proportional to the product $\beta \textrm{Kn}$ in dimensionless variables, are included in its derivation.
The dispersion relation is, however, still only adequate for describing scales that are both much longer than the mean-free-path of ion collisions (the fluid limit) and much shorter than the scale height of the atmosphere considered (the local limit). The modes of interest therefore need to fulfill Equation (\ref{eq:fluid_limit}).

In the resulting dimensionless variables the gradients in the temperature and the mean
molecular weight enter as $\dT$ and $\dmu$, making it easier to compare the results of this paper with previous work \citep{Pes13}.

\section{Stability Properties}
\label{sec:stability_properties}
The stability criterion for a stratified \emph{collisional} atmosphere is known
as the Schwarzschild criterion \citep{1958ses..book.....S}. According to this criterion, the plasma is stable
if the entropy increases with height, $z$, i.e., if
\be
\f{ds}{dz} > 0 \ .
\en
If the atmosphere is stratified in temperature and composition,
the criterion determining the stability of the atmosphere becomes
\be
\f{d\ln T/\mu}{d\ln P} < \f{2}{5} \label{eq:stab_criterion} \ .
\en
This is the Ledoux criterion known
from stellar convection theory \citep{1947ApJ...105..305L}.
Isothermal atmospheres with
\be
\f{d\ln\mu}{d\ln P} > - \f{2}{5} \ \label{eq:stab_only_mu},
\en
are therefore stable according to the Ledoux criterion. On the other hand, atmospheres with a uniform
composition need to fulfill
\be
\f{d\ln T}{d\ln P} < \f{2}{5} \ \label{eq:stab_only_T} \ .
\en
If Equation (\ref{eq:stab_criterion}) is not fulfilled a fluid element that is perturbed upwards (downwards)
will expand (contract) and continue to rise  (sink). We will refer to this type of
instability as gravity modes.

Atmospheres that satisfy the Ledoux criterion for stability (which assumes that the plasma is collisional)
are seen to be unstable when transport processes are anisotropic in a weakly collisional plasma.
When anisotropic heat conduction is taken into account, isothermal atmospheres with $ -2/5 < \dmu < 0$
are unstable regardless of the magnetic field inclination with respect to gravity. When anisotropic particle diffusion is considered even atmospheres with $\dmu > 0$ can become unstable.

The analysis carried out in \cite{Pes13}
shows that there are a host of instabilities that can feed off temperature and composition gradients
(see their Figures 2 and 4 for an overview of their results).
Here, we focus our attention on the instabilities that have the dominant growth rates for the cluster
model of \cite{Pen09} in the regime in which heat conduction is fast with respect to the dynamical
timescale, i.e., $\omc \gg \omdy$.  For convenience, we summarize here some of the most relevant
features of these instabilities:
\begin{enumerate}
\item The Magneto-thermo-compositional Instability (MTCI) has its fastest growth rate when the magnetic field is perpendicular to the direction of gravity.
In the limit of a weak magnetic field the MTCI stability criterion is
\be
\f{d\ln\mu/T}{d\ln P} > 0 \quad \textrm{if } b_x \neq 0 \ . \label{eq:crit_MTCI}
\en
As pointed out in \citet{Pes13}, this criterion for stability is not affected by anisotropic particle diffusion. This feature of the MTCI is explained in further detail in Section \ref{sec:diffusion_of_Helium_fordmu_m1}.
\item The Heat- and Particle-flux-driven Buoyancy Instability (HPBI) has its fastest growth rate when the magnetic field is parallel to the direction of gravity.
If we ignore particle diffusion and magnetic field tension, the HPBI stability criterion is
\be
\f{d\ln\mu T}{d\ln P} > 0 \quad \textrm{if } b_z \neq 0, \, \omd \ll \omdy \ .
\label{eq:crit_HPBI}
\en
Note that, even if this criterion is fulfilled, overstable modes might be present, see Equation (57) in \cite{Pes13}.
\item The diffusive HPBI, which depends on anisotropic diffusion of particles, has its fastest growth rate when the magnetic field is parallel to the direction of gravity. The diffusive HPBI ($\omd \neq 0$) has a stability criterion which is qualitatively different from the non-diffusive HPBI ($\omd=0$). The criterion for stability for the diffusive HPBI is
\be
\f{d\ln T/\mu}{d\ln P} > 0 \quad \textrm{if } b_z \neq 0, \, \omd \gg \omdy \ .
\label{eq:crit_D-HPBI}
\en
\item
A type of instability driven by anisotropic diffusion of Helium, which we refer to as diffusion modes, also depend on $\omd \neq 0$. Diffusion modes have their fastest growth rate when the magnetic field is parallel to the direction of gravity. The stability criterion for diffusion modes is\footnote{There is a typo in the text below Equation (63) in \cite{Pes13} where the stability criterion is missing an absolute value sign acting on the left hand side of the inequality. This typo does not affect any conclusions or figures in their paper.}
\be
\left| \f{d\ln T}{d\ln P} \right |> \left|\f{d\ln \mu}{d\ln P} \right|  \quad \textrm{if } b_z \neq 0, \, \omd \neq 0 \ .
\label{eq:crit_diffusion-modes}
\en

\end{enumerate}

When the mean molecular weight is constant, Equation (\ref{eq:crit_MTCI}) reduces to the stability criterion for the MTI and Equations (\ref{eq:crit_HPBI}) and (\ref{eq:crit_D-HPBI}) both reduce to the stability criterion for the HBI. These instabilities, driven by thermal gradients in weakly collisional, homogeneous plasmas, have been studied in great detail \citep{Bal00,Bal01,Par05,Par07,Par08,Par08_MTI,Qua08,Kun11, Kun12, Lat12,Par12}.

Before considering the instabilities that are present when the temperature and
composition gradients are those  obtained from current sedimentation models,
we consider a series of simpler cases in order to build our intuition.

\section{Application to isothermal atmospheres }
\label{sec:isothermal}

In order to shed light on the instabilities that are driven by composition gradients we focus our attention on the case of
isothermal atmospheres in which the mean molecular weight increases with height. For these atmospheres, the stability criteria for the
HPBI and the MTCI, Equations (\ref{eq:crit_HPBI}) and (\ref{eq:crit_MTCI}),  both reduce to $\dmu > 0$.  In the following we analyze simple magnetic field geometries and assume, for simplicity, that the magnetic field strength is
negligible and thus $\oma = 0$. In this section, we calculate the growth rates associated with
axisymmetric modes as a function of the wavenumbers $k_x$ and $k_z$. Here, $k_x$ is the wavenumber perpendicular to gravity and $k_z$ is the wavenumber parallel to gravity. The latter corresponds to the radial direction in the ICM. We relate our results to the findings of \cite{Kun11}
who analyzed the MTI and the HBI in detail.

\subsection{Isothermal Atmospheres with No Particle Diffusion}

We solve the dispersion relation, Equation (\ref{eq:disp_relation}), for an isothermal atmosphere
with a gradient in composition in the limit where diffusion of particles is neglected ($\omd = 0$).

\subsubsection{Magnetic Field Perpendicular to Gravity}
\begin{figure}[t]
\centering
\includegraphics{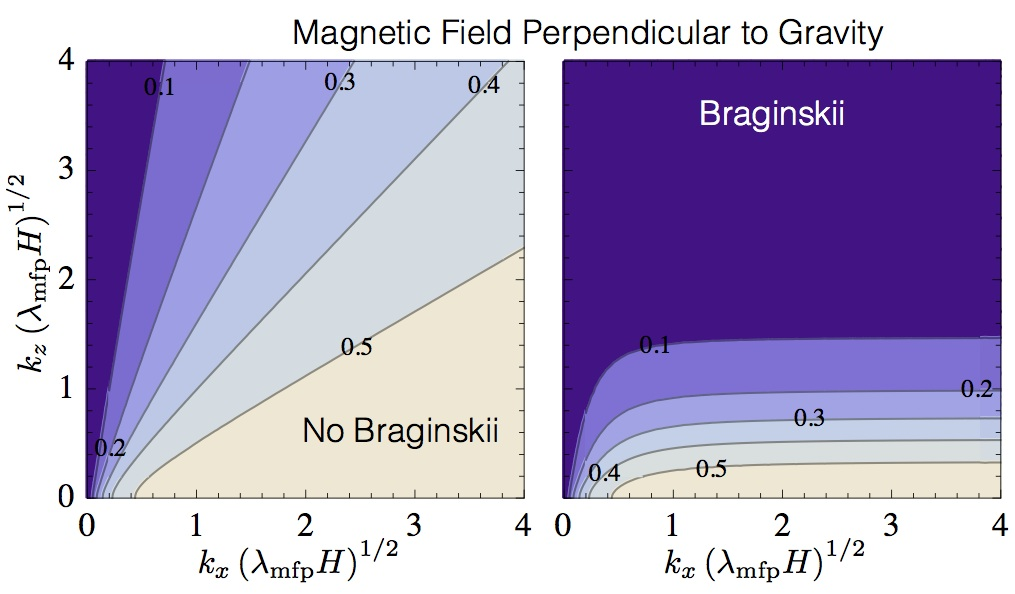}
\caption{Contour plots of $\sigma/\omdy$ for the MTCI with $\dmu = -1/3 $ without (left) and with Braginskii viscosity (right). Because of the similarity between the MTI and the MTCI these figures are similar to Figure 5 in \cite{Kun11}.}
\label{fig_MTCI}
\end{figure}
We start out by considering the configuration where the MTCI is
maximally unstable, namely a horizontal magnetic field, i.e.,
 $b_x = 1$. We consider an atmosphere with $\dmu = -1/3$. This atmosphere is stable
according to the Ledoux criterion, Equation~(\ref{eq:stab_only_mu}), which
means that it would be stable if the plasma were collisional. It is however unstable according to the MTCI criterion
that applies in the weakly collisional regime.

There is an interesting correspondence between the MTI and MTCI, which is useful in
order to make connections with previous results. In order to illustrate this, let us ignore
particle diffusion of He ($\omd = 0$).  In this case, the dispersion relation given by Equation
(\ref{eq:disp_relation}) only depends on the gradients in temperature and mean molecular
weight through the combination $d\ln (T/ \mu) /d\ln P$. This means that the dispersion relation
for the MTCI at constant temperature with $\dmu = -1/3$, is identical  to the  dispersion relation for the MTI with $\dT= 1 /3$.
The crucial difference is of course that in the former case the instabilities are driven by the temperature
gradient, whereas in the latter case they are driven by the composition gradient.

The correspondence between MTI and MTCI is illustrated in Figure \ref{fig_MTCI}, where we show the growth rate of
unstable modes when $\dmu = -1/3 $ and obtain a similar result as \cite{Kun11} did for
the MTI with $\dT= 1/3 $.
In the left panel of Figure \ref{fig_MTCI}, Braginskii viscosity is not included and the
maximum growth rate has $k_z = 0$. The maximum growth rate of $\sigma/\omdy = 0.5$ is confined to a wedge in wavenumber space with $k_z \lesssim 0.5 k_x$. In the right panel of Figure \ref{fig_MTCI}, Braginskii viscosity is included and
the growth rate of $\sigma/\omdy = 0.5$ is now confined to a  thin band with $\kz \lesssim 0.3$.  We observe that the growth rates are only significant when $\kz \ll \kx$. This preference for parallel wavenumbers ($k_{\rm max} \approx k_\para$, where $k_{\rm max}$ is the wavenumber for $\sigma_{\rm max}$) is thoroughly investigated by \cite{Kun11}. Due to the identical dispersion relations for the MTI and the MTCI at constant temperature, we therefore refer
to Equations (62)  (without Braginskii viscosity) and (64)  (with Braginskii viscosity) in \cite{Kun11} for approximate limits on the magnitude of $k_\perp$ above which the growth rates become negligible.

\begin{figure}
\includegraphics{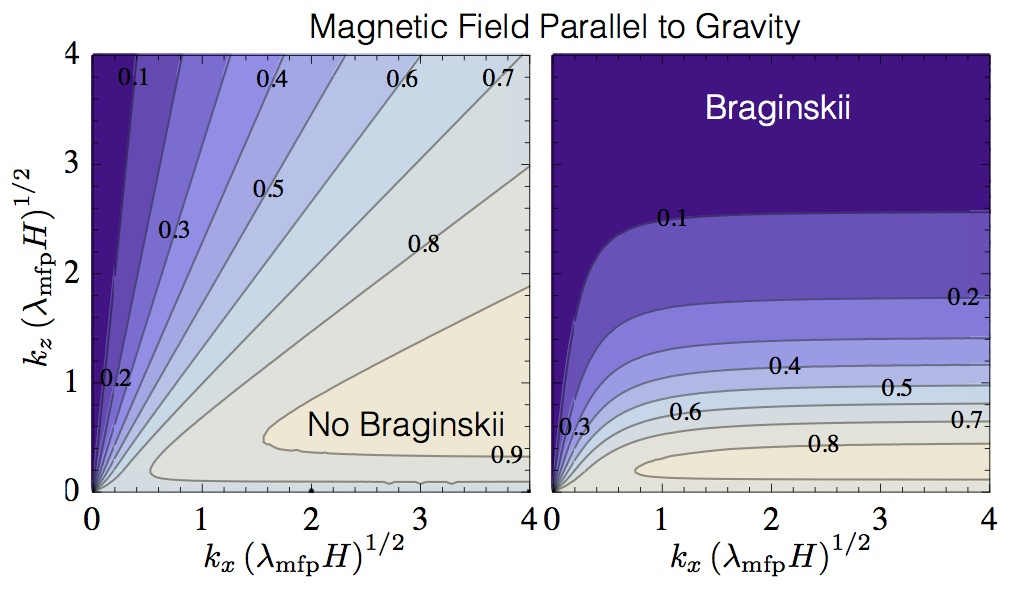}
\caption{Contour plots of $\sigma/\omdy$ for the HPBI with $\dmu = -1 $ without (left) and with Braginskii viscosity (right). The qualitative behavior is similar to the HBI but the maximum growth rates are located at smaller $\kz$. Gravity modes are seen at low $\kz$.}
\label{fig_HPBI}
\end{figure}

\subsubsection{Magnetic Field Parallel to Gravity}

Next, we consider the case of $b_z = 1$, i.e., a vertical magnetic field, where the HPBI is
maximally unstable. The growth rate for the HPBI as a function of wavenumber is shown in
Figure \ref{fig_HPBI} for the case of $\dmu = -1 $.

In the left panel of Figure \ref{fig_HPBI}, Braginskii viscosity is not included and the
large growth rates are confined to $k_z \lesssim k_x$.
In the right panel of Figure \ref{fig_HPBI}, Braginskii viscosity is included and the preference
for $k_\perp \gg k_\para$ is increased. The growth rate of $\sigma /\omdy = 0.8 $ is now
confined to $0.05 \lesssim \kz \lesssim 0.3$.
We conclude that the HPBI favors wavenumbers that have a large perpendicular component
and that  the available wavenumber space is a narrow band
with $k_{\perp} \gg k_\parallel$ when Braginskii viscosity is included in the analysis.

Figure \ref{fig_HPBI} looks remarkably similar to the corresponding figure for
the HBI (with $\dT = -1$) presented in \cite{Kun11} but they are not identical
at low $\kz$.
Even though the HBI and HPBI both have the property that $\sigma = 0$ for $k_\para = 0$ we observe that $\sigma/\omdy \approx 0.7$ along the line of $k_\para = 0$ in both the left and right panels of Figure \ref{fig_HPBI}.
The explanation is that gravity modes are unstable for different signs of the
logarithmic derivatives of $T$ and $\mu$, as seen in Equations (\ref{eq:stab_only_mu})
and (\ref{eq:stab_only_T}). The growth rate of gravity modes for $\dmu = -1$
at these wavenumbers has the value $\sigma/\omdy \approx 0.7$.
The reason for gravity modes at low $\kz$ is that heat conduction is too slow to drive the  HPBI when $k_{\para}$ is small. The gravity modes do not depend on heat conduction and they are therefore dominant in this slow conduction limit.

The gravity modes are not seen at high $\kz$ because they are damped by Braginskii viscosity.
Even though the HPBI is also damped by Braginskii viscosity, the HPBI turns out to have a higher growth rate than the gravity modes at high $\kz$. Gravity modes are present even in the absence of anisotropic transport, as illustrated in the left panel of Figure \ref{fig:pure_gravity_mode}. The damping of gravity modes by Braginskii viscosity is demonstrated in the right panel of Figure \ref{fig:pure_gravity_mode}.

 An important conclusion in \cite{Kun11} is that the local mode analysis
 for the HBI is not strictly valid when Braginskii viscosity is taken into account
 because the largest growth rates are obtained for $k_\para < H^{-1}$, implying that Equation
 (\ref{eq:fluid_limit}) is not satisfied.  The HPBI has its maximum growth rate at even longer  wavelengths than for the HBI and we therefore reach a similar conclusion for the HPBI at constant temperature as \cite{Kun11} did for the HBI. A quasi-global
 model has been developed for the HBI by \cite{Lat12}. This kind of approach can also be
 generalized to develop quasi-global models including composition gradients.

\begin{figure}
\centering
\includegraphics{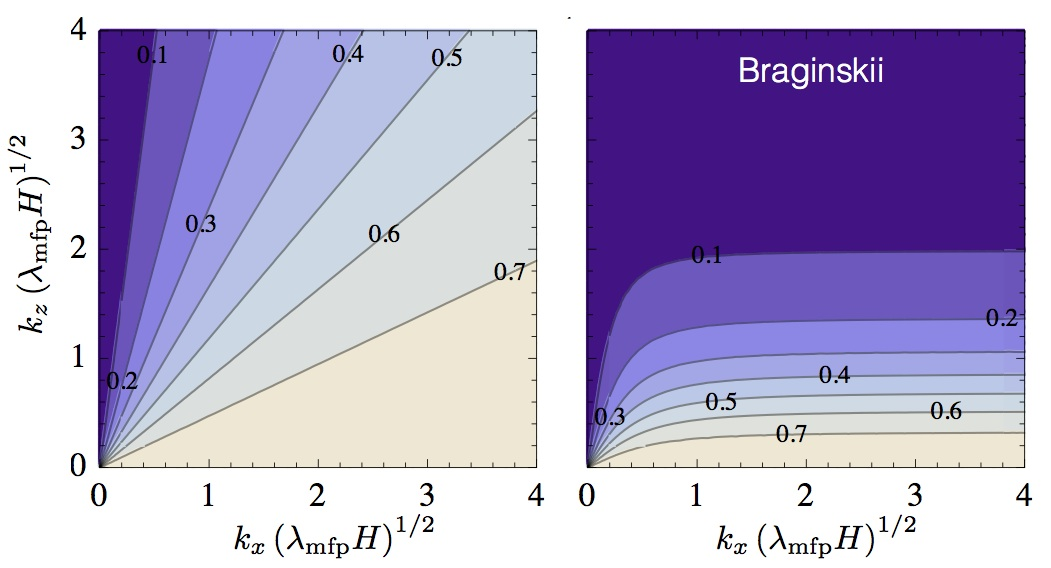}
\caption{Contour plots of $\sigma/\omdy$ for gravity modes with $\dmu = -1$ which appear even without anisotropic transport. \emph{Left:}   Without Braginskii viscosity and the growth rate of gravity modes is independent of the magnetic field inclination. \emph{Right:} With Braginskii viscosity and the gravity modes with $\kz\gtrsim2$ have negligible growth rate due to damping by Braginskii viscosity. The damping depends on the magnetic field inclination which is taken to be $\theta = 90^\circ$.}
\label{fig:pure_gravity_mode}
\end{figure}

\begin{figure*}[t]
\centering
\includegraphics{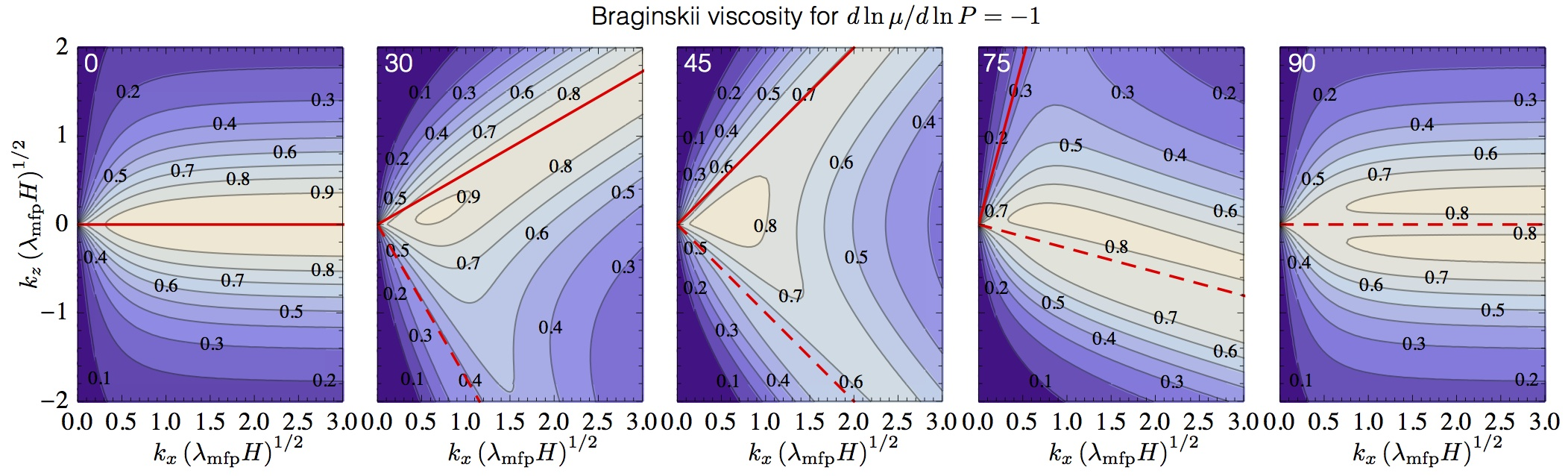}
\caption{Contour plots of $\sigma/\omdy$ for
an isothermal atmosphere
with $\dmu = -1$, with Braginskii viscosity. The red solid line indicates
$k = k_{\parallel}$ and the red dashed line indicates
$k_{\parallel} = 0$. The MTCI has its maximum growth rate for $k = k_{\parallel}$ and
the HPBI has its maximum growth rate with a small parallel component ($k \gg k_{\parallel}$). The inclination of the magnetic field is written in degrees at the top, left corner in each panel.
}
\label{fig:dlnmu_is_-1_angles_neg}
\end{figure*}

\begin{figure*}[t]
\centering
\includegraphics{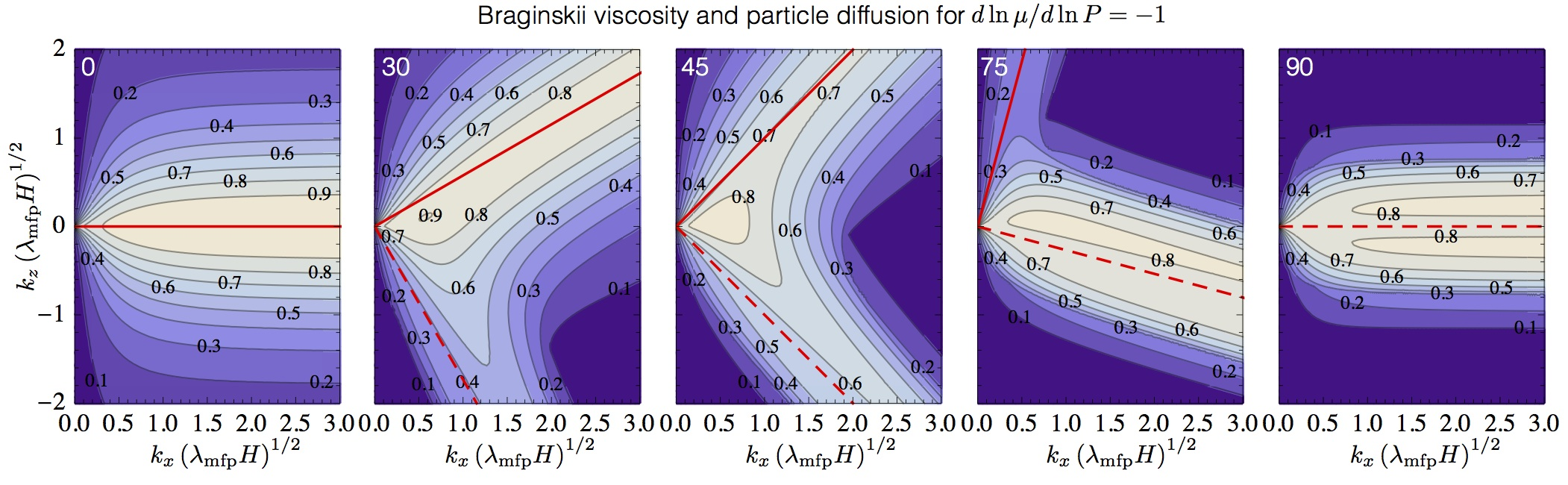}
\newline
\caption{
Contour plots of $\sigma/\omdy$ for
an isothermal atmosphere
with $\dmu = -1$, with Braginskii viscosity and particle diffusion. The red solid line indicates
$k = k_{\parallel}$ and the red dashed line indicates
$k_{\parallel} = 0$. The inclination of the magnetic field is written in degrees at the top, left corner in each panel.
From left to right: The MTCI at constant temperature is weakened
as the magnetic field becomes more vertical. It has its fastest growing modes along $k = k_\para$. The slow diffusion modes and the diffusive HPBI are
maximally unstable when the magnetic field is vertical. They are visible
from the second figure and onwards. In the last figure the field is entirely vertical and
the MTCI is stabilized. The instabilities that remain are the slow diffusion modes, the diffusive HPBI and the gravity modes.
}
\label{fig:dlnmu_is_-1_angles_D}
\end{figure*}

\subsubsection{More General Magnetic Field Geometries}

In this section, we explore the consequences of the presence of a magnetic field which is inclined at an angle $\theta$
with respect to the horizontal. The components of $\b$ are thus given by $b_x = \cos\theta$ and
$b_z = \sin \theta$.
In the previous two sections, and in agreement with  \cite{Kun11}, we showed that Braginskii viscosity
can play a significant role in the growth of modes driven by composition gradients.
Therefore, we include both anisotropic heat conduction and Braginskii viscosity in our analysis.

We consider an atmosphere with $\dmu = -1$
which is maximally unstable to the MTCI when $\theta = 0^\circ$
and to the HPBI when $\theta = 90^\circ$.
Figure \ref{fig:dlnmu_is_-1_angles_neg} shows the unstable modes that emerge as
the inclination of the magnetic field is varied, increasing from $\theta = 0^\circ$ in
the leftmost panel to $\theta = 90^\circ$
in the rightmost panel. The directions of $\bb{k} = \bb{k}_\para$ and $\bb{k} = \bb{k}_\perp$  are
indicated with a red solid line and a red dashed line, respectively.
In the previous sections we argued that the MTCI has its maximum growth rate
for wavenumbers with $k_\para \gg k_\perp$ while the HPBI has its maximum growth
rate for $k_\perp \gg k_\para$.  This provides an intuitive way to interpret Figure \ref{fig:dlnmu_is_-1_angles_neg}
which illustrates that the isothermal atmosphere is unstable regardless of the magnetic field inclination, $\theta$.

The results displayed in Figure \ref{fig:dlnmu_is_-1_angles_neg}
can be analyzed further with the insights gained earlier in this section.
Only the MTCI is unstable in the first panel ($\theta = 0 \, \rm ^\circ$) and
the most unstable wavenumbers lie in a band along $k_z = k_\perp = 0$. In the second panel,
this unstable band is
rotated to lie along $\theta = 30^\circ$, which is the angle of $\bb{k}_\para$ with respect
to the horizontal. At the same time, a new unstable band has appeared
in the direction $\bb{k}=\bb{k}_\perp$.
This is the HPBI which prefers $k_\perp \gg k_\para$. We note again that both the MTCI and the HPBI have zero growth rate along $k_\para = 0$ (the dashed line) and so the growth rates seen along this line must be due to gravity modes.

In the third panel ($\theta = 45^\circ$), both unstable bands have rotated by another 15 degrees. The maximum growth
rate of the HPBI (MTCI) unstable band has increased (decreased) to $\sigma/\omdy = 0.6$ ($\sigma/\omdy = 0.7$).
The maximum growth rate is found in a region around $k_z = 0$ and it difficult to associate
this wavenumber with a specific instability.
In the fourth panel ($\theta = 75^\circ$), the  maximum growth
rate of the HPBI unstable band has increased even further and it is now
larger than the maximum growth rate of the MTCI whose growth rate
has decreased down to $\sigma/\omdy = 0.3$. In the final  panel ($\theta = 90^\circ$),
the MTCI
is completely stabilized and only the modes associated with the HPBI remain.

\begin{figure*}[t]
\centering
\includegraphics{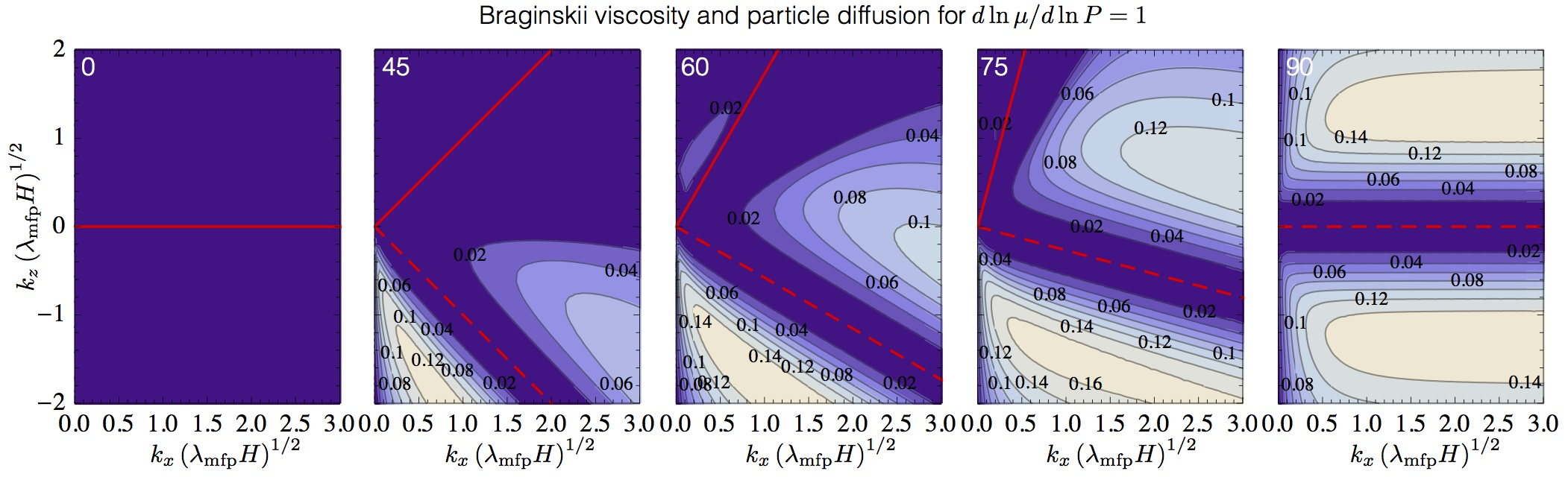}
\caption{The diffusion modes for $\dmu = 1$. This instability is driven by diffusion of He and is
only unstable when $D \neq 0$ and $b_z 	\neq 0$. The diffusion mode with $\dmu = 1$ operates on a smaller length scale than the diffusion mode with $\dmu = -1$.\label{fig:dlnmu_is_1_angles} }
\end{figure*}

\pagebreak

\subsection{Diffusion of Helium}
In this section, we consider the stability properties of a
weakly collisional, weakly magnetized, binary plasma
with Braginskii viscosity and anisotropic diffusion of particles. We focus our attention
on isothermal atmospheres that are stratified in composition.

\subsubsection{An Atmosphere with $\dmu = -1$}
\label{sec:diffusion_of_Helium_fordmu_m1}

We start out by considering an atmosphere with $\dmu = -1$ which is unstable to gravity modes. When the effect of anisotropic diffusion of He is ignored, this atmosphere is generally unstable to both the MTCI and the HPBI, as shown in the previous section. In this section diffusion of particles is included in the analysis.

The anisotropic diffusion of He enables a number of new processes \citep{Pes13}.
First and foremost, a new type of instabilities, termed diffusion modes, appear. These modes
only exist due to anisotropic diffusion of Helium and their growth rate increase with the value of the  diffusion coefficient, $D$.
Second, a new type of instability, termed the diffusive HPBI \citep{Pes13} appears in place of the HPBI when $\omd \gg \omdy$.

The stability criterion for the  MTCI is unaffected by the presence of particle diffusion,
as seen in Equations (60) and (68) in \cite{Pes13}.
The fact that the stability criterion of the MTCI is unaffected
can
be understood intuitively by considering a fluid parcel moving upwards in a gravitational potential while being connected to its previous surroundings by a magnetic field line. The vertical displacement of the parcel gives rise to an expansion of the parcel. In the the absence of heat transfer to the parcel this expansion would lead to a decrease in the temperature. Due to anisotropic heat conduction, however, the magnetic field line is effectively an isotherm. The parcel is therefore heated from below, rendering it unstable. This mechanism for instability is the same as for the MTI but with the mean molecular weight playing the role of the temperature in the background atmosphere. The mean molecular weight is initially constant along the field line and it is unaffected by expansions or contractions of the fluid parcel. The vertical displacement does therefore  not give rise to any anisotropic particle diffusion along the field line and we conclude that the MTCI should be largely unaffected by $D \neq 0$.

The new features enabled by particle diffusion are illustrated
in Figure~\ref{fig:dlnmu_is_-1_angles_D}, which only differs from Figure~\ref{fig:dlnmu_is_-1_angles_neg}
in that particles are able to diffuse along magnetic field lines, i.e., $\omd = k_\parallel^2 \lambda_{\rm mfp} H \omdy/4$.
The first and leftmost panels of Figure  \ref{fig:dlnmu_is_-1_angles_neg} and \ref{fig:dlnmu_is_-1_angles_D}
are identical because the MTCI is mostly unaffected by particle diffusion, as explained above.
The second panel shows the MTCI unstable band along $\bb{k}_\para$
and the diffusion modes along $\bb{k}_\perp$. In the third panel, where the magnetic field is
inclined at $45\deg$,  there is a region of stability between the
two bands of unstable modes. This is in stark contrast with the
third panel of Figure  \ref{fig:dlnmu_is_-1_angles_neg} where the corresponding region has significant growth rates.
The fifth and rightmost panel of Figure \ref{fig:dlnmu_is_-1_angles_D} can be roughly divided into two unstable bands: An inner band with a growth rate of $\sigma/\omdy = 0.7$ and an outer band with a maximum growth rate of $\sigma/\omdy = 0.2$. The inner band has the same growth rate as the gravity modes seen in the fifth panel of Figure \ref{fig:dlnmu_is_-1_angles_neg} and the second panel of Figure \ref{fig_HPBI}. The maximum growth rate of $\sigma /\omdy= 0.8$ is confined to a small area of wavenumber space. Furthermore, we see that a large region of wavenumber space is stable when particle diffusion is included. We conclude that the instabilities have an even stronger tendency to prefer $k_\perp \gg k_\para$ when particle diffusion is included.

\subsubsection{An Atmosphere with $\dmu = 1$}

Next, we consider an atmosphere with $\dmu = 1$. This atmosphere
is only unstable when anisotropic diffusion of particles is taken into
account. Furthermore, the unstable modes require that the magnetic field has
a vertical component, i.e. $b_z \neq 0$.
The growth rates of the diffusion modes are shown in Figure \ref{fig:dlnmu_is_1_angles}.
These diffusion modes have a preference for $k_\perp \gg k_\para$ but they have
zero growth rate if $k_\para = 0$. Interestingly, they grow on a smaller length scale than the
diffusion modes found for $\dmu < 0$. This type of atmosphere (isothermal with the mean
molecular weight decreasing with height) is relevant in the context of the boundary between the intermediate and the outer ICM
in the model of \cite{Pen09} that we discuss next.

\section{Applications to sedimentation models  }
\label{sec:sedimentation}

Having gained some insight into the various instabilities that can be triggered by the
presence of a composition gradient in an isothermal environment, we now address
the stability properties of more realistic scenarios, relevant to the conditions expected
in the ICM. In order to accomplish this, we consider one of the models for Helium sedimentation introduced in \cite{Pen09}.
Before we present the analysis of the stability of a cluster model in which
Helium has sedimented efficiently, we provide a brief summary of the assumptions
and procedure involved in deriving these models.

\subsection{Spherically Symmetric Helium Sedimentation Models}

In the He sedimentation model of \cite{Pen09} the plasma is assumed to be in hydrostatic equilibrium
in the gravitational potential that is mainly due to dark matter. The composition is initially uniform with $c = 0.25$ ($\mu = 0.59$) at all radii as given by the primordial abundance of Helium.
The temperature of the cluster has a radial dependence that is motivated by observations and it is fixed in time. This amounts to assuming that heating and cooling are balanced at all radii. This assumption is
 also made, for instance,  in \cite{2010MNRAS.401.1360S}.
Furthermore, because the stellar mass content of a cluster is smaller than the total
Helium mass, enrichment from galaxies can be ignored \citep{Mar07}.

Given this initial setup, the Burgers' equations \citep{1969fecg.book.....B,Tho94} are solved for each ion species of the plasma assumed to consist of Hydrogen and Helium ions, as well as electrons.
The diffusion velocity of the Helium ions is found by assuming that the gravitational force on the Helium ions is balanced by the force due to electric fields, the gradient in partial pressure and resistance due to collisions with Hydrogen ions. The result is a slow gravitational settling of Helium ions toward the core and the development of a non-uniform composition profile.
This change in the composition of the gas takes the cluster out of hydrostatic equilibrium. This is caused by the change in the pressure which depends on the mean molecular weight, $\mu$, see Equation (\ref{eq:eos}). Burgers' equations only describe the relative motion of the species and so a momentum equation for the bulk flow of the gas needs to be solved in order to describe the restoration of hydrostatic equilibrium.
The radial distribution of Hydrogen and Helium is evolved by repeating these steps
over cosmological timescales.

\subsection{Evolving the Composition in Sedimentation Models}

A brief summary of the calculations involved in the models described in detail in
\cite{Pen09} can be outlined as follows.

The total density distribution (gas + dark matter) of the cluster is given
by the Navarro-Frenk-White profile \citep{0004-637X-490-2-493}
\be
\rho_{\rm tot}(r) = \f{\rho_s}{r/r_s (1+r/r_s)^2} \ ,
\en
where $\rho_s$ is a normalization constant and $r_s$ is a characteristic scale. The total mass, $M(r)$,
enclosed within the radius, $r$, can be found by integrating the density distribution. This yields
\be
M(r) = 4\pi \rho_s r_s^3 \left[ \ln(1+r/r_s) - \f{r/r_s}{1+r/r_s} \right] \ ,
\en
from which one can find the gravitational acceleration at a distance $r$
\be
g(r) = \f{G M(r)}{r^2} \  , \label{eq:g_acc}
\en
where $G$ is the gravitational constant.

As per convention,
$r_{500}$ ($r_{2500}$) is defined to be the radius inside of which the mean density is 500 (2500) times the critical density of the universe. The value of $r_s$ is chosen such that $r_{s}=0.25 r_{500}$.
The model presented here has $r_{500} = 1.63$ Mpc, $r_{2500} = 0.75$ Mpc and a total cluster mass, $M(r_{500}) = 1.24\times 10^{15} \;M_\odot$ where
$M_\odot$ is the solar mass. The ratio of the mass of the ICM to the total mass of the cluster is assumed to be 0.15 at $r_{500}$.
The temperature profile is given by
\be
\f{T(r)}{T_0} = \f{ (r/0.045r_s)^{1.9}+  0.45}{(r/0.045r_s)^{1.9}+  1} \f{1.216}{\left[1+(r/0.6r_s)^2\right]^{0.45}}  , \,
\label{eq:T_cluster}
\en
where $T_0 = 10$ keV. The parameters used in the model and the functional dependence of $T(r)$ are motivated by a Chandra sample of 13 nearby, relaxed galaxy clusters \citep{Vik06}. The density, $\rho$, and pressure, $P$, of the gas is found by solving the equation of hydrostatic equilibrium
\be
\f{dP}{dr} = -\rho g(r) \ ,
\en
where
the gravitational potential is given by Equation (\ref{eq:g_acc}) and the pressure is related to the density by Equation (\ref{eq:eos}).

Burgers' equations, namely the continuity and momentum equations for each species, $s$, are given by \citep{1969fecg.book.....B,Tho94}
\be
  \frac{\partial n_s}{\partial t}+ \frac{1}{r^2}\frac{\partial (r^2n_s u_s)}{\partial r} &=&0 \ ,
  \\
  \frac{\partial P_s}{\partial r} + n_s A_s m_{\rm H} g - n_s Z_s eE &=& \sum_t K_{st} (w_t - w_s)  \ .
\en
 where $n_s$ is the number density, $u_s$ is the velocity, and $w_s$ is the velocity of species $s$ relative to the bulk
velocity of the fluid, $u$. The mass and charge numbers of ion species $s$ are given by $A_s$ and $Z_s$. The electric field is given by $E$
and the resistance coefficients are given by
\be
 K_{st}  = f_{\rm B}^{-1} \frac{4 \sqrt{2 \pi}}{3}
 \frac{e^4Z_s^2Z_t^2 m_{st}^{1/2}}{(k_{\rm B} T)^{3/2}} n_s n_t \ln \Lambda_{st}  \ ,
 \en
where $m_{st}$ is the reduced mass and $\ln \Lambda_{st}$ is the Coulomb logarithm of species $s$ and $t$, which we set to 40 (see more details in Appendix \ref{sec:transport_properties}).
The parameter $f_{\rm B}^{-1}$ is the magnetic suppression factor that regulates
the slow down of the sedimentation process envisioned to arise as a result of tangled magnetic fields.
The profiles shown in Figure~\ref{fig:cluster_Tandmu} have been obtained by setting $f_{\rm B}=1$
and thus ignore this effect.

\begin{figure}
\centering
\includegraphics[trim = 7 0 0 0, width=0.51\textwidth]{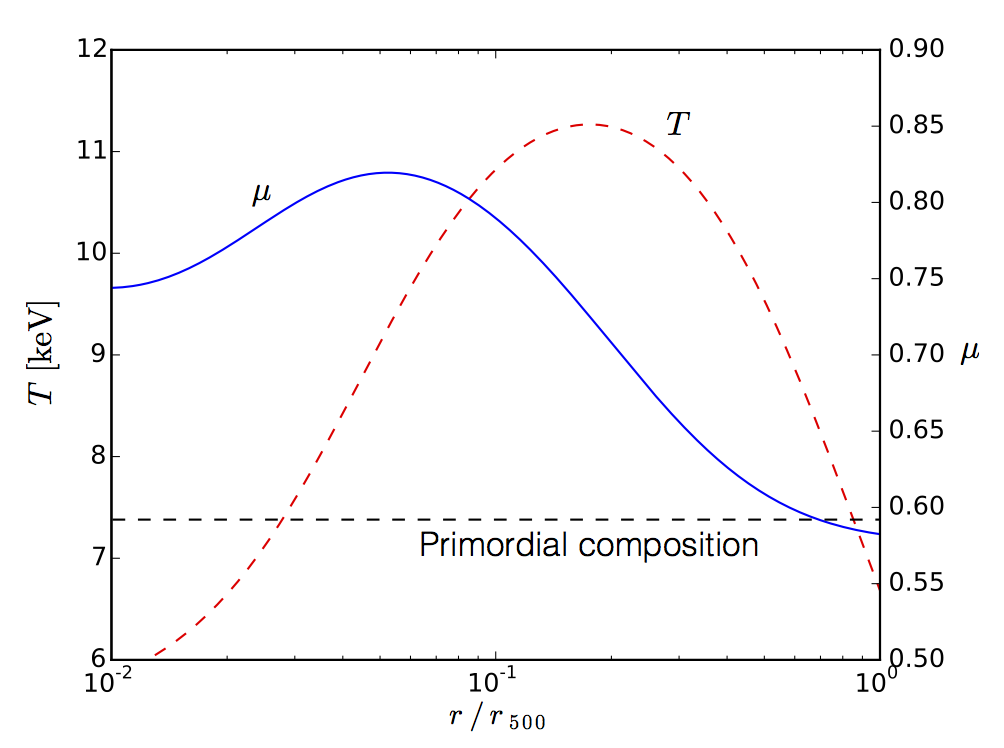}
\caption{The mean molecular weight profile (blue line) in a 11 Gyr-old galaxy cluster and the temperature profile (red dashed line) used in the model of \citet{Pen09}. The dashed black line indicates the primordial
composition at $t = 0$ Gyr which is $\mu = 0.59$.}
\label{fig:cluster_Tandmu}
\end{figure}

Burgers' equations are solved along with the momentum equation for the bulk motion of the gas
\be
\rho \f{du}{dt} = -\D{P}{r}-\rho g(r) \ ,
\en
and the distribution of elements is found as a function of time.

We show the results from a calculation using this method in Figure \ref{fig:cluster_Tandmu} which was produced by rerunning the code\footnote{The authors of \cite{Pen09} kindly provided us with a copy of the original Fortran code used for their paper.} developed by \cite{Pen09}. In this figure, the mean molecular weight profile for a 11 Gyr-old cluster is shown along with the temperature profile used for the calculation.
A simple explanation of the peak in the mean molecular weight is that the resistance coefficient depends strongly on temperature.  This means that He sedimentation will tend to be most effective where the temperature is high. The dashed black line shows the initial (primordial) composition of the plasma.

Efficient sedimentation in the ICM can lead to biases in the estimates of key parameters of clusters if the sedimentation is not taken into account in the data analysis. The specific model described here would lead to biases of
6\% in the total mass and gas mass at $r = r_{2500}$ if a homogeneous plasma is assumed. This would create a bias of 12\% in the gas mass fraction of the cluster and a bias of around 20\% in the estimate for the Hubble constant (see Figure 4 in \citealt{Pen09}). In the next section we discuss how the composition profile inferred from the model could be unstable at all radii due to plasma instabilities.

\begin{figure}
\centering
\includegraphics[trim=0 0 0 -8,  width=0.45\textwidth]{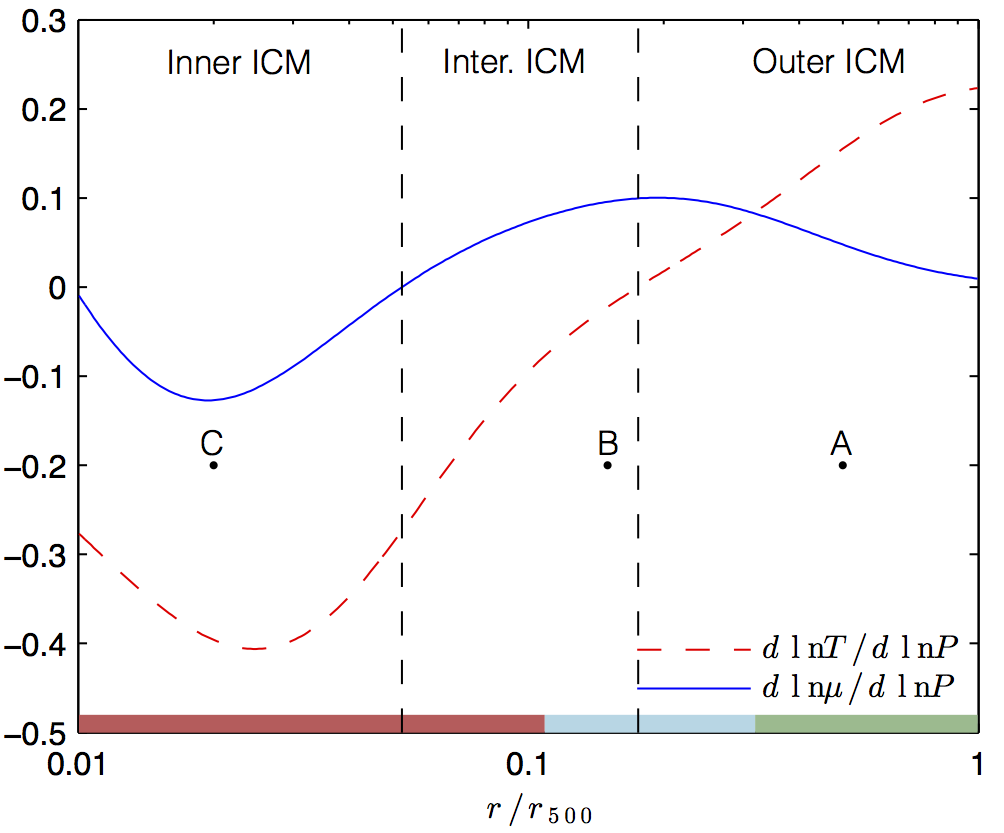}
\caption{The gradients $\dT$ (red dashed line) and $\dmu$ (blue line) as a function of radius in the model of \cite{Pen09} at $t = 11$ Gyr. The ICM is divided into inner ICM, intermediate ICM and outer ICM.
The color bar
illustrates which instabilities could be triggered,
 see Figure \ref{fig:cluster_parametric} for the color coding.
The labels $A$, $B$ and $C$ indicate the radii
used to produce Figure \ref{fig_Outer}, \ref{fig_Inter} and \ref{fig_Inner}.}\label{fig:cluster_gradients}
\end{figure}

\subsection{Stability Analysis of Helium Sedimentation Models}
\label{sec:Stability_analysis_of_Peng09}

The model presented in \cite{Pen09} provides estimates for the derivatives
$\dT$ and $\dmu$ as a function of radius. This is illustrated in Figure \ref{fig:cluster_gradients},  where we have divided the ICM into three regions. The inner ICM
which extends from $r/r_{500} = 0.01$ to the radius where $\dmu = 0$ ($r/r_{500} = 0.05$) and
the outer ICM which extends from the radius where $\dT = 0$ ($r/r_{500} = 0.18$) to
the radius $r/r_{500} = 1$. The intermediate ICM is defined to be the region in between the previous two.

Before delving into details, we can provide a qualitative idea about which parts of the ICM are prone to the different types of
instabilities. In order to do this, we consider  the numerical values of the gradients in temperature
and composition in the context of the stability diagrams introduced in \cite{Pes13} (see Figure 2 and 3
in their paper). The comparison is facilitated by using a parametric plot in the $(\dmu,\,\dT)$ plane,
see Figure \ref{fig:cluster_parametric}. The colored sections in this figure indicate the regions of
parameter space which are subject to the different types of instabilities discussed in Section
\ref{sec:stability_properties}. The extent of these regions is also indicated with
color bars at the bottom of Figure \ref{fig:cluster_gradients}. The red color bar
indicates the region where the diffusive HPBI could be active and the blue bar indicates the region
where the diffusive HPBI and the diffusion modes could be active. Finally, the region highlighted by
the green bar is unstable to the MTCI and the conduction modes.\footnote{The conduction modes are important in the slow conduction limit, $\omc \ll \omdy$, see \citet{Pes13}. } These conclusions require, of course,
that the magnetic field geometry allows for the various instabilities to be triggered.

\begin{figure}[t]
\centering
\includegraphics{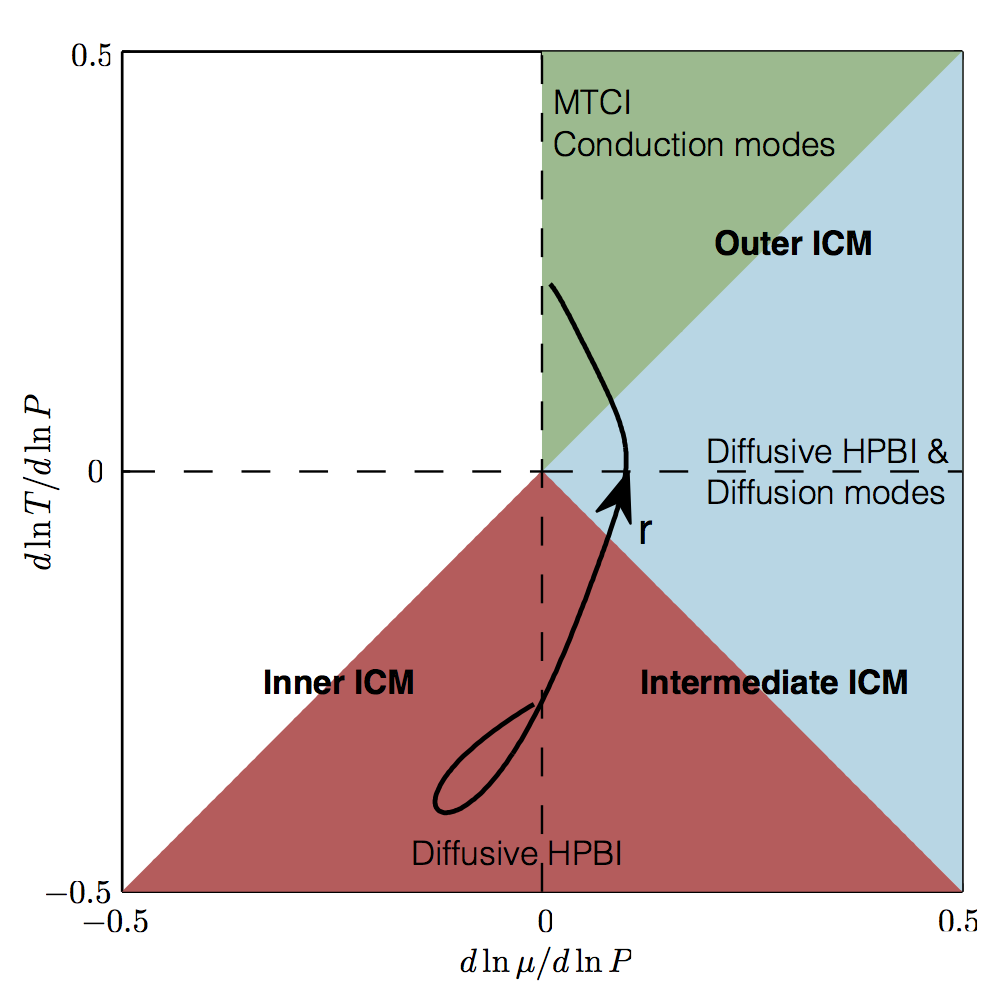}
\caption{Parametric plot in the $(\dmu,\dT)$ plane as a function of $r/r_{500}$ (black line with arrow indicating the direction of increasing radial distance). Red is unstable to the diffusive HPBI. Blue is unstable to the diffusive HPBI and the diffusion modes. Green is unstable to the MTCI and the conduction modes. Comparing this figure with the stability diagrams of \cite{Pes13} leads to the color bars identifying the different regions in Figure \ref{fig:cluster_gradients}.}
\label{fig:cluster_parametric}
\end{figure}

The stability criteria derived in  \cite{Pes13} assume that magnetic tension is negligible.
In order to assess whether this effect could be important,
we need a model for the magnetic field strength in the ICM. We can estimate the plasma $\beta$
as a function of cluster radius,  $\beta(r)$, for the model of \cite{Pen09}, by using
\be
B(r) = B_0 \left(\f{n_{\rm e}(r)}{n_{\rm e}(0)} \right)^{\eta} \,, \label{eq:B(r)}
\en
where $n_e$ is the electron number density, $B_0 = 4.7$ $\mu$G, and $\eta = 0.5$,
as found for the Coma cluster in  \cite {Bon10}. The quantitative results therefore depend
on this choice while the qualitative results should not.
Using this model, we illustrate in Figure \ref{fig:cluster_validity} the potential role that magnetic tension could play,
especially in the inner parts of the ICM.
The assumptions made in \cite{Pes13} are valid within region 4 in Figure \ref{fig:cluster_validity}. The dispersion relation derived in this paper extends the validity of the analysis to also include regions 2 and 5 where magnetic tension is important.
For the sake of completeness, we recall that the fluid approximation breaks down in region 1 and 3 and that the local approximation in the linear analysis breaks down in region 6 and 7.

In what follows, we discuss  the growth rates and the characteristic distances on which the various instabilities
could operate in the cluster model of \cite{Pen09}  at $t = 11$ Gyr. Using the radial profiles provided by the model, we can calculate the values of the needed parameters ($\dT$, $\dmu$, $\omdy$, $H$, $\lambda_{\rm{mfp}}$, Kn, $\chi_\para$, $\nu_{\para}$, $D$) as a function of radius. In addition to the model of \cite{Pen09} we consider a magnetic field given by Equation (\ref{eq:B(r)}) to estimate $\beta$.

In order to assess the influence of the sedimentation of Helium on the stability properties of the ICM we solve the dispersion relation for the sedimentation model of \cite{Pen09} at both $t = 0$ Gyr and $t = 11$ Gyr.
We include anisotropic heat conduction, Helium diffusion,
 Braginskii viscosity and a finite $\beta$ in the calculations in Section \ref{sec:Outer}, \ref{sec:Inter} and \ref{sec:Inner}.
The effect of Braginskii viscosity is to damp perturbations with short perpendicular
wavelengths.
The effect of the magnetic tension is to stabilize modes with short parallel wavelengths and, in general, to inhibit the growth rates. The role of magnetic tension in decreasing the maximum growth rate is investigated in Section \ref{sec:discussion}.
In the following we consider each of the three regions of the ICM defined in Figure \ref{fig:cluster_gradients} separately.

\begin{figure}[t]
\centering
\includegraphics{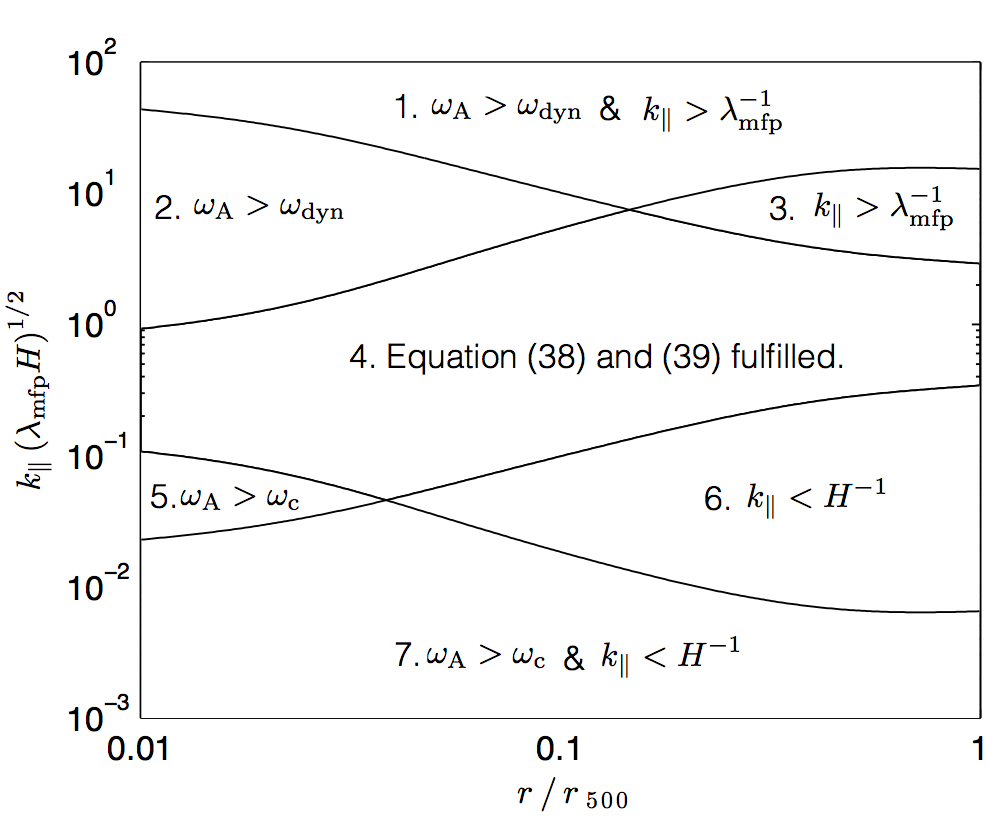}
\caption{Validity of the local linear analysis as a function of radius.  The results of \cite{Pes13} are valid within region 4. The effects of magnetic tension should be accounted for in regions 2 and 5. The fluid approximation breaks down in region 1 and 3 and the local approximation breaks down in region 6 and 7.}
\label{fig:cluster_validity}
\end{figure}

\subsection{Outer ICM}
\label{sec:Outer}
In the model of \cite{Pen09} at $t = 11$ Gyr, both the temperature and the mean
molecular weight decrease with radial distance in the outer ICM, as illustrated in Figure \ref{fig:cluster_Tandmu}.
In the inner part
of the outer ICM we have $\dmu > \dT$, see Figure \ref{fig:cluster_gradients}, and this makes this region unstable to diffusion
driven modes and the diffusive HPBI at short parallel wavelengths and to the
diffusion modes at long parallel wavelengths.
In the outer part of the outer
ICM we have $\dmu < \dT$ which makes this region unstable to the MTCI
and to conduction modes at long parallel wavelengths. The model at $t = 0$ Gyr, before Helium has had time to sediment, is unstable to the MTI. We note that these conclusions depend on the magnetic field geometry.

As an illustration, we consider a magnetic field inclined at $\theta = 45^\circ$ at a specific radial distance, $r/r_{500} = 0.5$,
indicated with a letter $A$ on Figure \ref{fig:cluster_gradients}. Using values evaluated at this location ($\dT = 0.16$ and $\dmu = 0.05$ at $t = 11$ Gyr and $\dT = 0.16$ and $\dmu = 0$ at $t = 0$ Gyr) we calculate the growth
rates, as shown in Figure \ref{fig_Outer}.
In the panel on the left (right) we show the growth rate as a function of wavenumber for the model at $t= 0$ Gyr ($t =11$ Gyr).
We observe that the maximum growth rate
is $\sigma \approx 0.87 \textrm{ Gyr}^{-1}$
without Helium sedimentation ($t = 0$ Gyr) and $\sigma \approx 0.75 \textrm{ Gyr}^{-1}$ with Helium sedimentation ($t=11$ Gyr)
such that the instabilities grow unstable on a timescale
of either 1.15 Gyr or 1.3 Gyr, respectively. The presence of Helium sedimentation is concluded to lead to a decrease in the growth rate by approximately 15\%, in agreement with the rough estimate in \cite{Pes13}.
Considering a characteristic scale $L = 2\pi/k$, the fastest growing mode corresponds to $(L_x ,\, L_z) = (1.0 ,\, 1.0)$ Mpc at $t = 0$ Gyr. This scale is slightly decreased to $(L_x ,\, L_z) = (0.9 ,\, 0.8)$ Mpc at $t = 11$ Gyr.
 The unstable modes found are describable by a fluid approach
(at this radial distance $\lambda_{\rm{mfp}} = 30$ kpc) but they are not strictly describable by a local linear analysis (at this radial distance $H = 0.35$ Mpc).
The value of $\omdy^{-1}$ at this distance is roughly 0.3 Gyr so the instability grows on a timescale a factor of a few larger than the dynamical timescale.

\begin{figure}[t]
\centering
\includegraphics{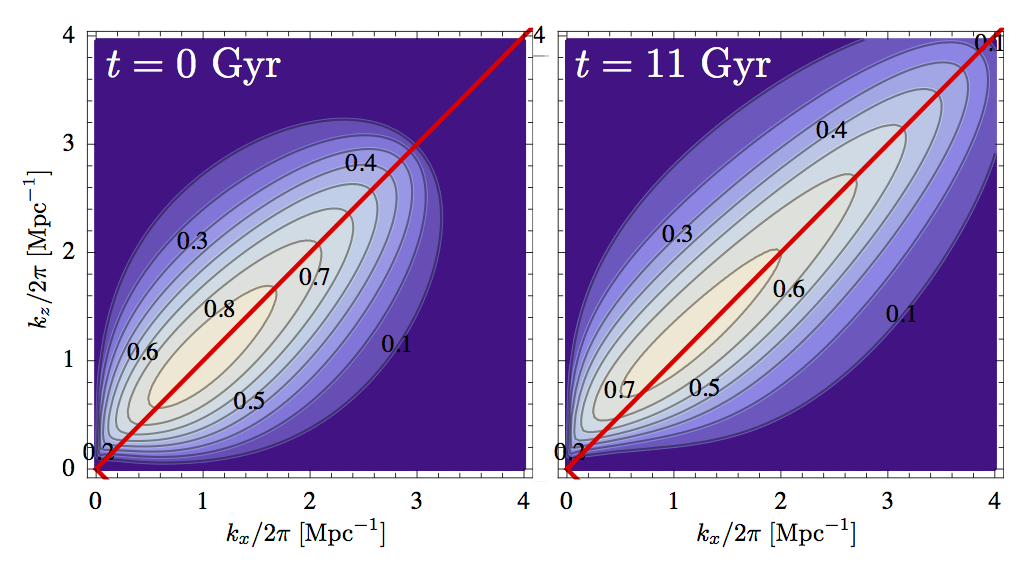}
\caption{
Contour plots of the growth rate in Gyr$^{-1}$ for the outer ICM (location C). This region is unstable to the MTI at $t= 0$ Gyr (left panel) or the MTCI and the conduction modes at $t = 11$ Gyr (right panel).
The most unstable modes are found at intermediate parallel wavenumbers because magnetic tension stabilizes modes with a high parallel wavenumber. The maximum growth rate is decreased by 15\% with respect to the homogeneous case.}
 \label{fig_Outer}
\end{figure}

\subsection{Intermediate ICM}
\label{sec:Inter}
According to the model at $t = 11$ Gyr, the temperature increases while the mean
molecular weight decreases with radial distance in the intermediate ICM.
The stability diagrams of \cite{Pes13} then reveal that the intermediate
ICM is unstable to the diffusive HPBI in the entire region. Furthermore,
the outer part of the intermediate ICM is unstable to the diffusion
modes.

For illustrative purposes, we consider the radial distance
indicated with a letter $B$ on Figure \ref{fig:cluster_gradients} which is
located at $r/r_{500} = 0.15$. At this location, $\dT = -0.02$ and $\dmu = 0.1$ at $t = 11$ Gyr
and so the diffusion modes and the diffusive HPBI are expected to be active.
In the absence of sedimentation this radial distance is unstable to the HBI ($\dT = -0.03$ and $\dmu = 0$ at $t = 0$ Gyr).
We consider the growth rates for the $t = 0$ Gyr cluster in
the left panel and the $t = 11$ Gyr in the right panel of Figure \ref{fig_Inter}. In this figure we have assumed $\theta = 75^\circ$.
 The intermediate region is stabilized by the gradient in composition at $t = 11$ Gyr if anisotropic particle diffusion is neglected ($D = 0$) but when anisotropic particle diffusion is taken into account ($D \neq 0$) the diffusive HPBI and diffusion modes could be active. This can understood from the criteria for stability for the HPBI, the diffusive HPBI and the diffusion modes,
 given by Equations (\ref{eq:crit_HPBI}), (\ref{eq:crit_D-HPBI}) and (\ref{eq:crit_diffusion-modes}), respectively. While Equation (\ref{eq:crit_HPBI}) is satisfied Equations (\ref{eq:crit_D-HPBI}) and (\ref{eq:crit_diffusion-modes}) are not. We conclude that even though diffusion modes only grow on a diffusive timescale, $\omd^{-1}$, they can be dominant if the instabilities that grow on a conduction timescale, $\omc^{-1}$, are not active. The diffusive HPBI, however, requires that $\omd \gg \omdy$, a requirement that is not fulfilled in the unstable region in Figure \ref{fig_Inter}. The diffusion modes are active regardless of whether $\omd \gg \omdy$ or $\omd \ll \omdy$, and so the growth rates present in Figure \ref{fig_Inter} are interpreted to be due to the diffusion modes.
The maximum growth rates are $\sigma \approx 0.28 \textrm{ Gyr}^{-1}$
at $t = 0$ Gyr and $\sigma \approx 0.59 \textrm{ Gyr}^{-1}$ at $t = 11$ Gyr corresponding to time scales of 3.6 Gyr and 1.7 Gyr, respectively. The maximum growth rate is increased by 110\% with respect to the homogeneous case. The most unstable scales are $(L_x ,\, L_z) = (0.32 ,\, 0.35)$ Mpc at $t = 0$ Gyr and $(L_x ,\, L_z) = (0.15 ,\, 0.15)$ Mpc at $t = 11$ Gyr. }

\begin{figure}[t]
\includegraphics{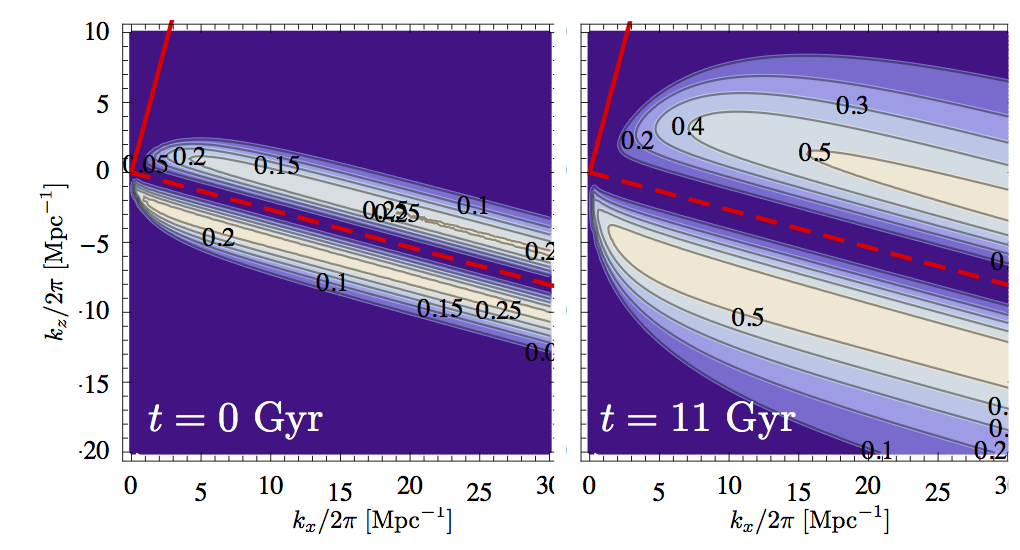}
\caption{
Contour plots of the growth rate in Gyr$^{-1}$ for the intermediate ICM (location B). This region is unstable to the HBI at $t= 0$ Gyr (left panel) or the diffusion modes and the diffusive HPBI at $t= 11$ Gyr (right panel). This region is only unstable at $t= 11$ Gyr if $D \neq 0$. The maximum growth rate is increased with 110\% with respect to the homogeneous case.}
\label{fig_Inter}
\centering
\end{figure}

\subsection{Inner ICM}
\label{sec:Inner}
In the inner ICM both the temperature and the mean
molecular weight increase with radial distance at $t = 11$ Gyr. This implies that this region is only unstable with respect to
the diffusive HPBI at $t = 11$ Gyr. At $t = 0$ Gyr, it is unstable to the HBI. The magnetic field strength increases towards the center of the ICM and
so we expect the magnetic tension to dampen the growth rates more severely
in the inner ICM.

\begin{figure}[t]
\centering
\includegraphics{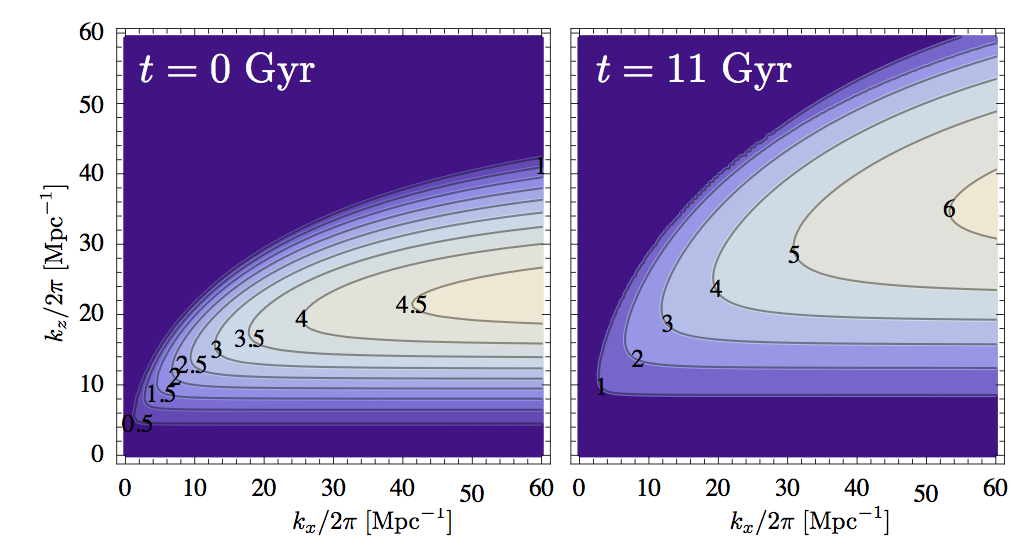}
\caption{
Contour plots of the growth rate in Gyr$^{-1}$ for the inner ICM (location A). This region is unstable to the HBI at $t = 0$ Gyr (left panel) and to the diffusive HPBI at $t = 11$ Gyr (right panel). In both cases, the maximum growth rate is significantly decreased and the most unstable modes are found at longer parallel wavelengths because of magnetic tension. The maximum growth rate is increased by 40\% with respect to the homogeneous case.}
\label{fig_Inner}
\end{figure}

We consider the radial distance
indicated with a letter $C$ on Figure \ref{fig:cluster_gradients}, which is
located at $r/r_{500} = 0.02$. At this radius, $\dT = -0.4$, $\dmu = -0.13$ at $t = 11$ Gyr and $\dT = -0.51$, $\dmu = 0$ at $t = 0$ Gyr.
Due to the low value of  $\beta \rm Kn \approx 2$, we expect magnetic tension to influence
the dynamics as highlighted in Figure \ref{fig_Inner}. We assume that $\theta = 90^\circ$ which is the maximally unstable configuration. In Figure \ref{fig_Inner}, the left
panel shows the growth rates at $t = 0$ Gyr and the right panel shows the growth rates at $t = 11$ Gyr.
Braginskii viscosity makes the HPBI have a preference for $k_\perp \gg k_\para$, as explained in Section \ref{sec:isothermal}. Magnetic tension also acts to inhibit the growth of modes with a high parallel wavenumber. Braginski viscosity and magnetic tension are therefore the reasons for the zero growth rates at high vertical wavenumbers, $k_z$.
The wavenumbers are not restricted in the $x$-direction (perpendicular to gravity) and so the fastest growth rates are attained for short distances in the
$x$-direction because heat conduction is effective on short distance scales.
The maximum growth rate is therefore found for $L_z \approx 25$ kpc and an even shorter length scale in the $x$-direction. The fluid limit is, however, only valid as long as $L_x \gg \lambda_{\rm{mfp}} \approx 40$ pc at this distance. The vertical length scale of the fastest growing mode should be much smaller than $H \approx 50$ kpc but this is not the case.
The maximum growth rates are $\sigma \approx 5 \textrm{ Gyr}^{-1}$ without sedimentation and $\sigma \approx 7.2 \textrm{ Gyr}^{-1}$ with sedimentation corresponding to time scales for growth of 0.20 Gyr and 0.14 Gyr, respectively. When sedimentation is present, we find that the maximum growth rate is increased by 40\% with respect to the homogeneous case.

\pagebreak

\subsection{Magnetic tension decreases the growth rates}
\begin{figure}[t]
\centering
\includegraphics{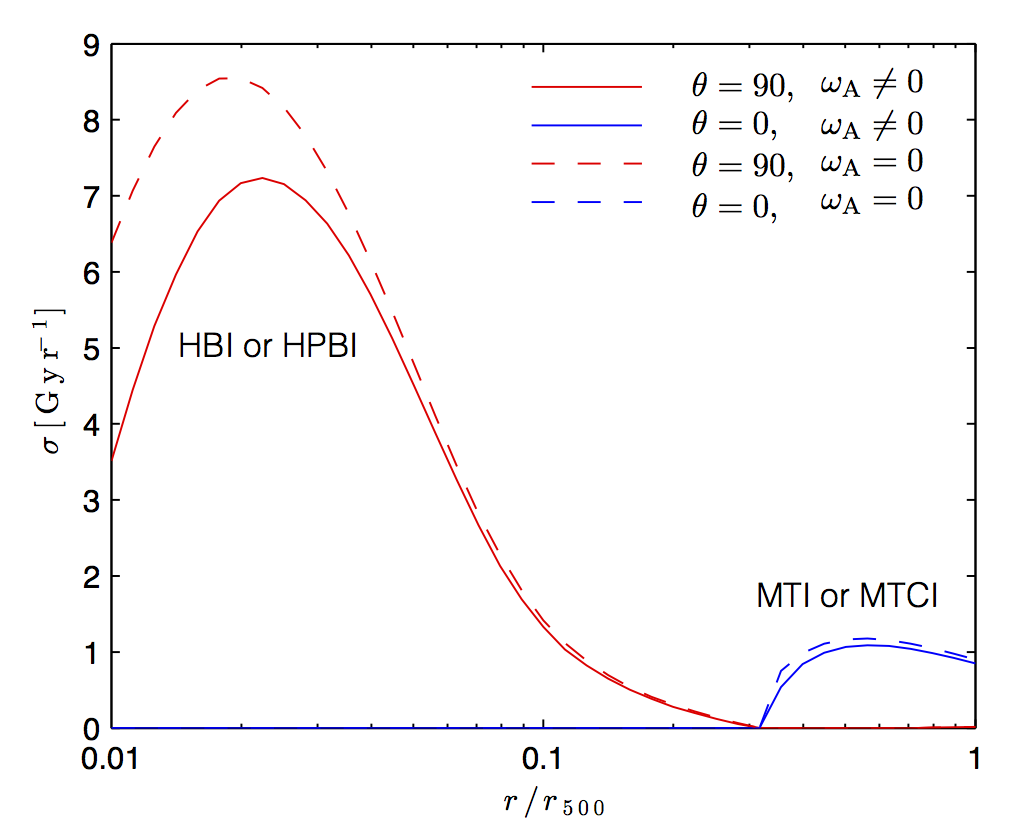}
\caption{The maximum growth rates as a function of radius in the cluster model of \cite{Pen09} at $t = 11$ Gyr in the limit where magnetic tension is neglected (dashed lines) and when it is taken into account (solid lines).}
\label{fig:tension_vs_no_tension}
\end{figure}

In order to assess how the effect of magnetic tension modifies the growth rates we compare the solutions we obtain when we set $\oma = 0$ with
those found when we set $\oma$ equal to the value found by combining the model of \cite{Pen09} at $t = 11$ Gyr with Equation (\ref{eq:B(r)}).
The maximum growth rates as a function of radius for a field with $\theta = 90^\circ$ and $\theta = 0^\circ$ inclination with respect to the direction of gravity are shown in Figure \ref{fig:tension_vs_no_tension}. The solid lines include $\omega_{\textrm{A}} \neq 0$ while the dashed lines are found by solving the $\omega_{\textrm{A}} = 0$ limit of the dispersion relation. We conclude that magnetic tension decreases the maximum growth rate at all radii but the effect is seen to be most significant in the inner cluster region \citep{Car02,Kun11,Pes13}.

\section{Discussion and Prospects}
\label{sec:discussion}
Understanding whether He sedimentation in galaxy clusters is efficient or whether it can be hindered
by tangled magnetic fields, turbulence, or mergers remains an open question in astrophysics.
Addressing this problem from first principles demands a better understanding of the processes
involved in  the weakly collisional, magnetized plasma constituting the ICM. As a first step in
this endeavor, we have taken a simple approach to gauge the importance of various dynamical
instabilities, related to the MTI and HBI, that can feed off temperature and composition gradients
\citep{Pes13} as expected from state-of-the-art sedimentation models \citep{Pen09}.

\begin{figure}[t]
\centering
\includegraphics{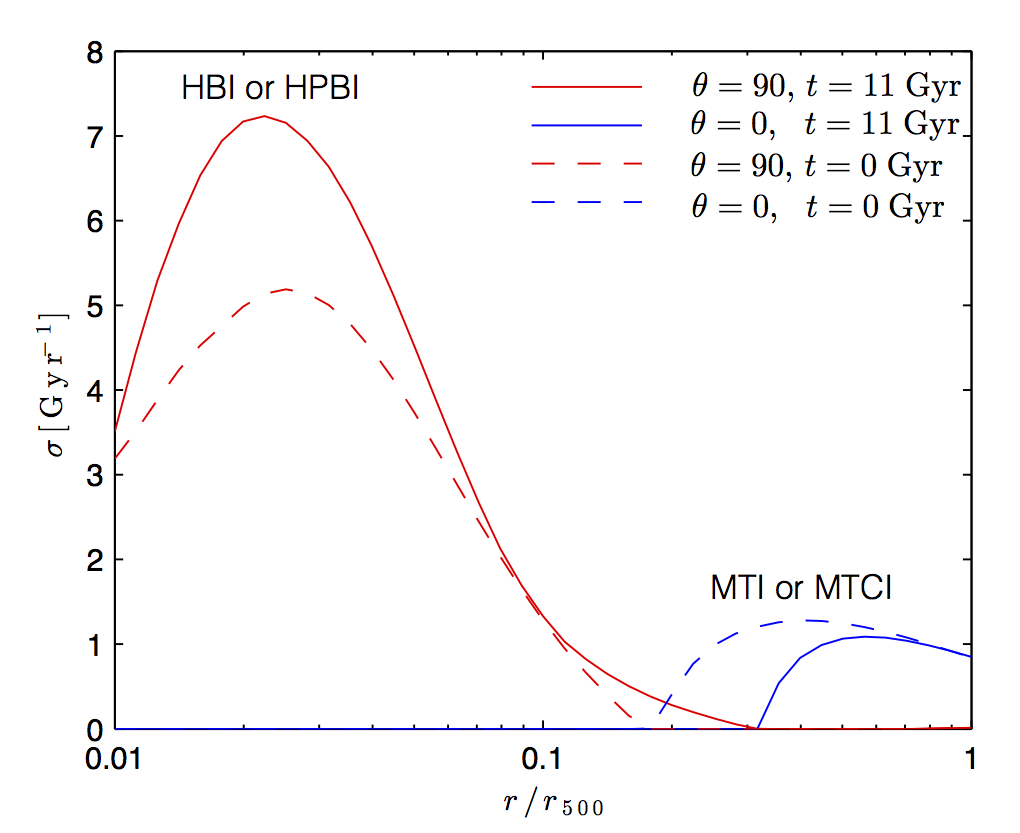}
\caption{The maximum growth rate as a function of radius in the cluster model of \cite{Pen09} at $t = 0$ (dashed) and $t = 11$ Gyr (solid) for a field with $\theta = 90^\circ$ (red) and $\theta = 0^\circ$ (blue) inclination with respect to the direction of gravity. The effects of a finite $\beta$ are included. The sedimentation increases (decreases) the theoretically predicted growth rates in the inner (outer) cluster.}
\label{fig:sedi_vs_no_sedi}
\end{figure}

We have shown that if a gradient in the composition of the ICM arises due to
Helium sedimentation, as modeled for example in \citet{Pen09}, this might not be a stable
equilibrium. We illustrated this by showing that, depending on the magnetic field orientation,
the radial profile of the sedimentation model is unstable, to different kinds of instabilities, at all radii. The instabilities are shown to grow on timescales that are short compared to the life-time of a typical cluster. Our findings are summarized in Figure \ref{fig:sedi_vs_no_sedi} where we show the maximum growth rate as a function of radius for both the homogeneous cluster model ($t = 0$ Gyr) and the cluster model with a gradient in composition ($t = 11$ Gyr) for a magnetic field that is either parallel or perpendicular to the direction of gravity. In this figure we find that, in accordance with \cite{Pes13}, Helium sedimentation can lead to an increase in the maximum growth rate in the inner cluster region but a decrease in the maximum growth rate in the outer cluster region.
The figure illustrates that the composition gradients, as inferred from sedimentation models which do not
fully account for the weakly collisional character of the environment,
are not necessarily robust even though the entropy increases with radius. This contrasts the arguments regarding the stability of composition gradients put forth in \citet{Mar07}, which predates the discovery of the HBI \citep{Qua08}.

The instabilities discussed in this paper could provide an efficient mechanism for diminishing the mean molecular weight gradient in the ICM by turbulently mixing the Helium content. Whether this is the case depends on how the instabilities saturate as well as the large scale dynamical processes that contribute to determining the global gradient in the mean molecular weight. There are several processes that could play a role in this regard at both small and large scales.
Understanding their influence will lead to a more realistic picture of the ICM dynamics.
We mention a few examples below.

The equations of kinetic MHD used in this paper do not incorporate the physics responsible for the composition gradients found in sedimentation models based on Burger's equations. They are therefore not able to self-consistently describe the coupling of magnetic fields to the sedimentation process. One possible route forward would be to extend the equations of kinetic MHD and take the sedimentation process into account
by following \cite{1990ApJ...360..267B}. This would allow us to describe the dynamical influence of the magnetic field at the cost of using a one-fluid model instead of the commonly used multifluid models. Even though an extension of the kinetic MHD framework would describe unmagnetized sedimentation less precisely than Burgers' equations \citep{Tho94}, this would be a step forward in our understanding of sedimentation processes in the ICM.

Our idealized model of the ICM consisted of a weakly collisional, plane-parallel
atmosphere in hydrostatic equilibrium. Real clusters are most likely not in perfect hydrostatic equilibrium as the ICM can be stirred by mergers and accretion. The ensuing turbulence can contribute with a significant fraction of the pressure support needed to counteract gravity \citep{2009ApJ...705.1129L,2014ApJ...792...25N}. The instabilities we have described could be influenced by such turbulence, as well as by the cosmological expansion over timescales comparable to the age of the Universe \citep{2011ApJ...740...81R}.

Another issue raised in this paper is that some of the fastest growing modes grow on scales that are not strictly local in height. This means that there is a need for a quasi-global theory as
developed in \cite{Lat12}
in order to correctly describe the linear dynamics of the weakly collisional medium. Other issues may affect
the plasma dynamics at small scales.
Very fast microscale instabilities, such as the firehose and mirror instabilities, could play a key role in the ICM \citep{schekochihin_fast_2006,Sch10,kunz_thermally_2011}. These instabilities are not correctly described in the framework of kinetic MHD \citep{schekochihin_plasma_2005}. This might not be a problem if the microinstabilities saturate in such a way that they drive the pressure anisotropy to marginal stability \citep{2008PhRvL.100h1301S,2011MNRAS.413....7R}. This is still an outstanding issue in the study of homogeneous plasmas. The microinstabilities are not a concern for the linear evolution of the MTI and the HBI but they are important for simulations of their nonlinear evolution \citep{Kun12}. We anticipate the need to deal with similar issues for simulations of the nonlinear evolution of the instabilities that are driven by gradients in composition.

\begin{figure*}[t]
\centering
\includegraphics{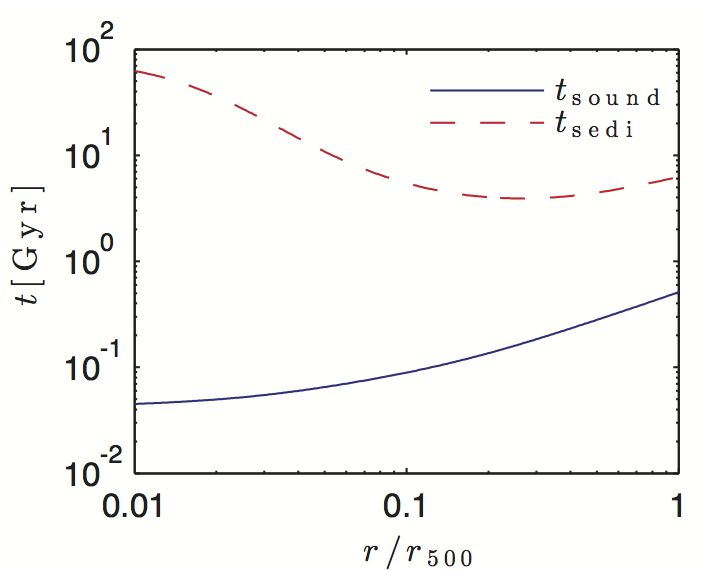}
\includegraphics[scale=0.5]{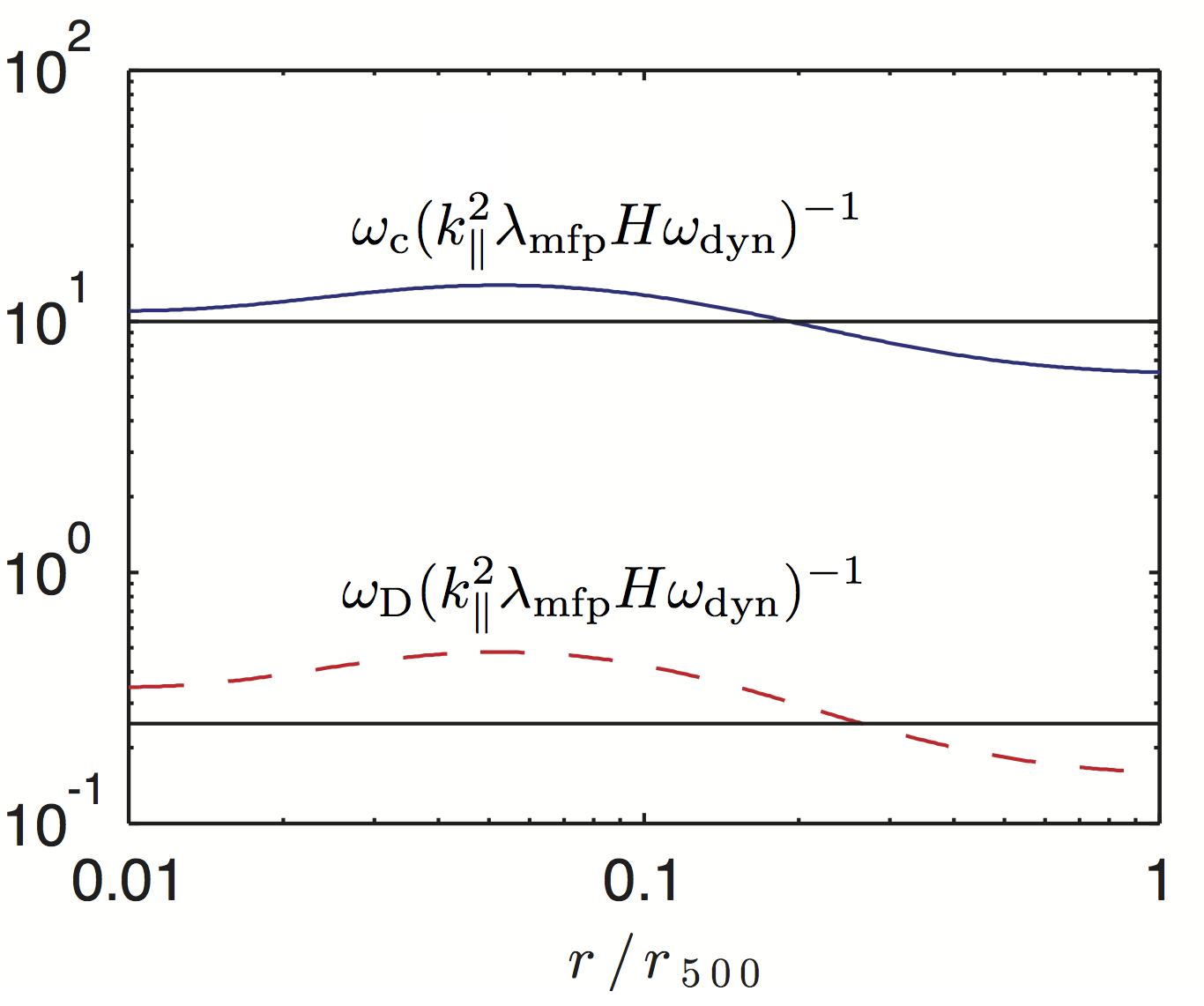}
\caption{\emph{Left:} The timescale for sedimentation across a scale height (red dashed) and the timescale for sound to cross a scale height (blue) as a function of radius in the model of \cite{Pen09}. \emph{Right:} The dimensionless values of $\omc$ (blue)
and $\omd$ (red dashed) are shown as a function of radius. The approximations given in Equations (\ref{eq:conduction_time}) and (\ref{eq:diffusion_time}) are indicated with solid horizontal lines.}
\label{fig:cluster_times}
\end{figure*}
\begin{figure*}
\centering
\includegraphics[width=\textwidth]{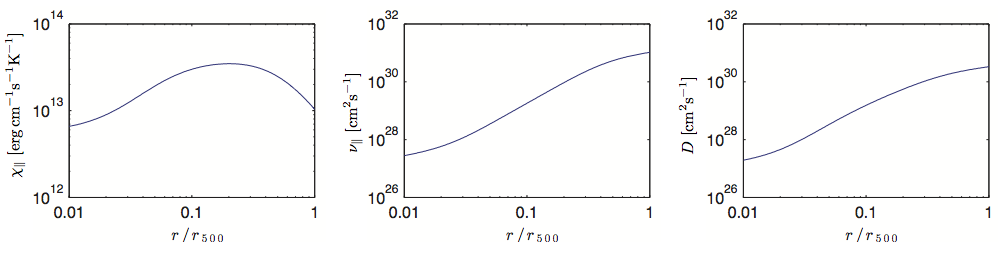}
\caption{The values of $\chi_\para$ (left), $\nu_\para$ (middle) and $D$ (right) as function of radius in the cluster model of \cite{Pen09} at $t = 11$ Gyr.}
\label{fig:cluster_coeffiecients}
\end{figure*}

\acknowledgements
We are grateful to Daisuke Nagai and Fang Peng for giving us a copy of their Fortran code that we used to reproduce the data in their sedimentation model. We are grateful
to the anonymous referee for a thoughtful and detailed report. The suggestion to make a more comprehensive comparison between a homogeneous and a heterogeneous ICM inspired us to produce several new figures and helped us to significantly improve the final version of this manuscript. We acknowledge useful discussions with Daisuke Nagai, Matthew Kunz, Prateek Sharma, Ellen Zweibel, and Ian Parrish during the \emph{$3^{\rm rd}$ ICM Theory and Computation Workshop} held at the Niels Bohr Institute in 2014. We also thank Sagar Chakraborty and Henrik Latter for valuable comments. The research leading to these results has received funding from the European Research Council under the European Union's Seventh Framework Programme (FP/2007-2013) under ERC grant agreement 306614. T. B. also acknowledges support provided by a L{\o}rup Scholar Stipend and M. E. P. also acknowledges support from the Young Investigator Programme of the Villum Foundation.

\appendix
\section{Characteristic time scales for sedimentation and anisotropic transport}

\subsection{Helium sedimentation}
\label{sec:sedi_timescales}
In this paper we have built on the stability analysis of \cite{Pes13} and applied these tools to the Helium profile provided by the sedimentation model of \cite{Pen09} in order to calculate the growth rates of instabilities that could be present in this model of the ICM. In our calculations we have assumed that the composition profiles evolve on timescales
that are longer than the characteristic timescales in which the instabilities operate. Within this framework,
we found that the relevant instabilities grow on timescales comparable to the dynamical timescale. We
show here that our approach is justified because the timescales involved in the sedimentation process are much longer than the dynamical timescale. In order to estimate the timescale for sedimentation, we use an approximation for the sedimentation velocity, $w_{\rm He}$, given by
\be
w_{\rm He} &\simeq & 80\usp{\rm km\,s^{-1}}
          \left(\frac{T}{10\usp{\rm keV}}\right)^{3/2}
\left(\frac{g}{10^{-7.5}\usp{\rm cm\,s^{-2}}}\right) \nonumber \\
&&\times
\left(\frac{n_{\rm H}}{10^{-3}{\rm cm^{-3}}}\right)^{-1} \ ,
\en
 in \cite{Pen09} for a single Helium ion immersed in a Hydrogen background. Here, $n_{\rm H}$ is the Hydrogen number density.
The separation of timescales is illustrated in Figure \ref{fig:cluster_times}
where we show the time for a He ion to sediment a distance of one scale height ($t_{\rm sedi} = H/w_{\rm He}$) along with the dynamical timescale ($t_{\rm sound} = H/v_{\rm th} = \omdy^{-1}$) as a function of radius in the cluster. We observe that the timescales differ by more than an order of magnitude, providing support to
our assumption.

\subsection{Heat conduction, Braginskii viscosity and particle diffusion}
\label{sec:time_scale_for_diffusion}
In order to estimate the timescale for particle diffusion we use Equations
(\ref{eq:chi_para}), (\ref{eq:eta_para}) and
(\ref{eq:Bahcael_dif}) to estimate the coefficients $\chi_\para$, $\nu_\para$ and $D$. We calculate the dimensionless values of $\omc$, $\omv$ and $\omd$ by scaling them with $k_{\para}^2 \lambda_{\rm mfp} H \omdy$
 where the frequencies $\omc$, $\omv$ and $\omd$ are defined in Equation~(\ref{eq:omega_c}).  The dimensionless values of $\omc$ and $\omd$ are  shown as a function of radius (using the model of \citealt{Pen09}) in the right panel of Figure \ref{fig:cluster_times}. From this figure we estimate that the diffusion timescale is roughly 40 times longer than the timescale for heat conduction, enabling us to estimate $\omd$ as in Equation~(\ref{eq:diffusion_time}). The dimensionless value of $\omv$ does not depend on any physical parameters and is therefore $3/2$ at all radii.

\section{Transport properties of a Hydrogen-Helium plasma}
\label{sec:transport_properties}
The procedure used to derive the kinetic MHD equations for a binary mixture is similar to the procedure used for
 a pure Hydrogen plasma \citep{Bra,1983bpp..conf....1K}.

The non-ideal transport coefficients for a plasma consisting of Hydrogen and Helium ions as well as electrons
are found by using the Krook
operator in the Vlasov-Landau-Maxwell equations
\be
\D{f_s}{t} + \bb{v} \bcdot \del f_s + \left[ \f{q_s}{m_s} \left(\bb{E}+\frac{\bb{v}\btimes\bb{B}}{c}\right) + \bb{g} \right]
 \bcdot \D{f_s}{\bb{v}} \label{eq:kinetic_full_v}
= C[f_s] \ .
\en
Here, $f_{s}$ is the one-particle phase-space distribution of species $s$ and $q_s$ ($m_s$) is the particle
charge (mass).
The Krook operator is given by  \citep{Sny97}
\be
C[f_s] = -\sum_t \nu_{st} \paren{f_s - F_{Mst}} \ , \label{eq:Krook}
\en
where the sum extends over all species and the equilibrium function, $F_{Mst}$, is given by
\be
F_{Mst} = \f{n_s}{\paren{2\pi T_s/m_s}^{3/2}} \exp \left[-\f{m_s}{2T_s} \paren{v_\parallel-u_{\para,t}}^2 -\f{m_s}{2T_s}v_\perp^2 \right]  \ . \qquad \label{eq:FMs}
\en
Here, the collision frequency between species $s$ and $t$
is given by
\be
\nu_{st} = \f{4 \sqrt{2\pi} }{3}  \f{n_t m_{st}^{1/2}e^4\paren{Z_sZ_t}^2}{m_s (\kb T)^{3/2} } \ln \Lambda_{st} \ ,
\en
where $\ln \Lambda_{st}$ is the Coulomb logarithm
and
\be
m_{st} = \f{m_s m_t}{m_s + m_t}\ ,
\en
is the reduced
mass. Furthermore, the mean velocity in the parallel direction of species $s$ is $u_{\para,s}$ and we assume that all species have the same temperature, $T_s = T$.

Following \cite{Pen09}, \cite{Sht10} we use $\ln \Lambda_{st} = 40$ which is a characteristic value for the ICM.
One can derive the kinetic MHD equations, given by Equations (\ref{eq:rho})-(\ref{eq:S}), by assuming that the distribution function is gyrotropic and calculating moments in velocity space of the Landau-Vlasov equation \citep{1983bpp..conf....1K}. If it is assumed that the distribution function is Gaussian
when calculating the moments $v_{\para}^4$, $v_\perp^4$, and $v_\para^2 v_\perp^2$, the equations are closed and one can show that (see, for instance, A.~A.~Schekochihin \& M.~W.~Kunz, in preparation) %
\be
\chi_\para = \f{5 }{2}\f{\kb}{\nu_{\rm e-e} + \nu_{\rm e-H} + \nu_{\rm e-He}} \f{P_e}{m_e} \ ,
\label{eq:chi_para}
\en
for the heat conductivity and
\be
\nu_\para &=&  \f{1}{\rho}\paren{\frac{n_{\rm He}}{\nu_{\rm He-H} + \nu_{\rm He-He}} +
\frac{n_{\rm H}}{\nu_{\rm H-H} + \nu_{\rm H-He} } } k_{\rm B} T \ , \,
\label{eq:eta_para}
\en
for the Braginskii viscosity. In these expressions the heat conduction due to ions
and the viscosity due to electrons is neglected. This approximation is good because
$m_{\rm H},\,m_{\rm He} \gg m_{\rm e}$.

The anisotropic diffusion coefficient due to a gradient in the composition was approximated by
\cite{1990ApJ...360..267B}. In terms of the ratio of the Helium density to the total gas density, $c$,
it can be expressed as \citep{Pes13}
\be
D=\frac{3}{16} \sqrt{\f{5m_{\rm H}}{2 \pi}}\frac{(k_BT)^{5/2}}{e^2 \rho\ln\Lambda_{\rm H-He}} \left[ \frac{4-c}{(2-c)(8-5c)}\right]\ . \label{eq:Bahcael_dif}
\en
Using the model of \cite{Pen09}, we show the transport coefficients as a function of radius in Figure \ref{fig:cluster_coeffiecients}.

\bibliography{Bibliography}

\end{document}